%% file: main.tex
\documentclass[sigconf,anonymous=false]{acmart}

\AtBeginDocument{%
  }

\acmConference[\color{white}]{}{}

\setcopyright{none}
\renewcommand\footnotetextcopyrightpermission[1]{} 
\input{utils} 

\usepackage{xspace}
\usepackage{enumitem}
\usepackage{caption} 
\usepackage{algorithmic}
\usepackage[noend]{algorithm2e}
\usepackage{url}
\usepackage{xcolor}

\SetAlCapSty{} 

\newcommand{\startsquarepar}{%
    \par\begingroup \parfillskip 0pt \relax}
\newcommand{\stopsquarepar}{%
    \par\endgroup}

\definecolor{mygray}{gray}{0.5}

\newcommand{\myitem}[1]{\vspace*{0.07in}\noindent\emph{\textbf{#1}}\enskip}

\newcommand{\myitemNoVspace}[1]{\noindent\emph{\textbf{#1}}\enskip}

\newcommand{\remove}[1]{}

\newcommand{\system}[0]{pForest\xspace} 

\newcommand{\Flow}[0]{F}
\newcommand{\Flowset}[0]{\mathcal{F}}
\newcommand{\RF}[0]{RF}
\newcommand{\rflabel}[1]{\text{label}(#1)}
\newcommand{\certainty}[1]{\text{certainty}(#1)}
\newcommand{\score}[1]{\operatorname{F_1}(#1)}

\newcommand{\thrS}[0]{\tau_s}
\newcommand{\thrC}[0]{\tau_c}
\newcommand{\classifier}[0]{\operatorname{C}}
\newcommand{\featurematrix}[0]{\operatorname{A}}
\newcommand{\features}[1]{\featurematrix(#1)}

\newcommand{\midistance}{\operatorname{D}}

\newcommand{\featurelist}[0]{\mathtt{F}}
\newcommand{\PPPP}[1]{P4$_{\text{#1}}$}
\newcommand{\cic}{CICIDS\xspace}
\newcommand{\unibs}{UNIBS\xspace}

\newcommand{\feati}[1]{\textcolor{gray}{\mathsf{F\,#1}}}

\newcommand{\ma}[0]{match$\kern 0.7pt$\&$\kern 0.7pt$action\xspace}

\begin{document}


\title{\vspace{-1em}pForest: In-Network Inference with Random Forests}


\author{Coralie Busse-Grawitz}
\affiliation{%
  \institution{ETH Z\"urich}
  \city{Z\"urich}
  \country{Switzerland}
  }
\email{bcoralie@ethz.ch}

\author{Roland Meier}
\affiliation{%
  \institution{ETH Z\"urich}
  \city{Z\"urich}
  \country{Switzerland}
  }
\email{meierrol@ethz.ch}

\author{Alexander Dietm\"uller}
\affiliation{%
  \institution{ETH Z\"urich}
  \city{Z\"urich}
  \country{Switzerland}
  }
\email{adietmue@ethz.ch}

\author{Tobias B\"uhler}
\affiliation{%
  \institution{ETH Z\"urich}
  \city{Z\"urich}
  \country{Switzerland}
  }
\email{buehlert@ethz.ch}

\author{Laurent Vanbever}
\affiliation{%
  \institution{ETH Z\"urich}
  \city{Z\"urich}
  \country{Switzerland}
  }
\email{lvanbever@ethz.ch}


\begin{abstract}
\input{abstract}
\end{abstract}

\maketitle


\input{sec_introduction}
\input{sec_model}
\input{sec_overview}

\input{sec_time_dependent_random_forests}

\input{sec_compilation_to_p4}

\input{sec_inference}
\input{sec_implementation}
\input{sec_evaluation}

\input{sec_discussion}
\input{sec_related}
\input{sec_conclusion}

\message{^^JLASTBODYPAGE \thepage^^J}

\bibliographystyle{ACM-Reference-Format}
\newpage
\bibliography{refs/refs_new}

\message{^^JLASTREFERENCESPAGE \thepage^^J}

\clearpage
\nobalance

\appendix
\input{app_training_alg}

\end{document}

%% file: utils.tex
\newcommand{\xref}[1]{\S\ref{#1}}

\newcommand{\pxref}[1]{(\xref{#1})}
\newcommand{\fref}[1]{Fig.~\ref{#1}}
\newcommand{\tref}[1]{Table~\ref{#1}}
\newcommand{\first}{\emph{(i)}\xspace}
\newcommand{\second}{\emph{(ii)}\xspace}
\newcommand{\third}{\emph{(iii)}\xspace}
\newcommand{\fourth}{\emph{(iv)}\xspace}
\newcommand{\fifth}{\emph{(v)}\xspace}
\newcommand{\sixth}{\emph{(vi)}\xspace}

\newcommand{\First}{\emph{(I)}\xspace}
\newcommand{\Second}{\emph{(II)}\xspace}

\newcommand{\ie}{i.e., \@}
\newcommand{\eg}{e.g., \@}

\newcommand{\cf}{cf. \@}

\newcommand{\etal}{et~al.\xspace}

\newcommand{\perc}{\,\%\xspace}

\newcommand{\bigO}[1]{\mathcal{O}\left(#1\right)}

\newenvironment{smallitemize}{%
\begin{itemize}[leftmargin=*]%
\setlength{\parsep}{0.3ex}%
\setlength{\leftmargin}{-0pt}%
\setlength{\listparindent}{-12ex}%
\setlength{\itemsep}{3pt}%
\setlength{\topsep}{0ex}%
\setlength{\parskip}{0ex}%
\setlength{\partopsep}{0ex}%
}{%
\end{itemize}
}%

\definecolor{darkgreen}{rgb}{0,0.5,0}
\definecolor{brown}{rgb}{0.7,0.3,0}
\definecolor{darkblue}{rgb}{0,0,0.5}
\definecolor{darkred}{rgb}{0.5,0,0}

\newcounter{fn1}
\newcounter{fn2}
\newcounter{fn3}
\newcounter{fn4}
\newcounter{fn5}

%% file: abstract.tex
When classifying network traffic, a key challenge is deciding when to perform
the classification, i.e., after how many packets. 
Too early, and the decision basis is too thin to classify a flow confidently; 
too late, and the tardy labeling delays crucial actions (e.g., shutting 
down an attack) and invests computational resources for too long (e.g., 
tracking and storing features). Moreover, the optimal decision timing varies
across flows. 

\startsquarepar
We present pForest, a system for 
``As Soon As Possible'' (ASAP)
in-network classification according to supervised machine learning models on
top of programmable data planes. pForest automatically classifies 
each flow 
as soon as its label is sufficiently established, 
not sooner, not later. A key challenge behind pForest is finding a strategy 
for dynamically adapting the features
and the classification logic during the lifetime of a flow. pForest solves
this problem by: (i) training random forest models tailored to different
phases of a flow; and (ii) dynamically switching between these models in real
time, on a per-packet basis. pForest models are tuned to fit the
constraints of programmable switches (e.g., no floating points, no loops,
and limited memory) while providing a high accuracy.
\stopsquarepar

We implemented a prototype of pForest in Python (training) and P4 (inference).
Our evaluation shows that pForest can classify traffic ASAP for hundreds of
thousands of flows, with a classification score that is on-par with software-based
solutions. 

%% file: sec_introduction.tex
\section{Introduction} \label{sec:introduction}

Classifying network traffic in real time is an online classification problem which consists
in predicting label(s) from a possibly unbounded stream of packets. Examples of
classification tasks abound in networking and include: 
detecting attacks (e.g., DDoS)~\cite{doshiMachineLearningDDoS2018}, 
distinguishing mice flows from elephant flows~\cite{zhangPHeavyPredictingHeavy2021},
identifying encrypted video streams~\cite{schusterBeautyBurstRemote2017},
predicting flow length~\cite{dukicAdvanceKnowledgeFlow2019},
website fingerprinting~\cite{hayesKfingerprintingRobustScalable2016},
device fingerprinting~\cite{formbyWhoControlYour2016,meidanProfilIoTMachineLearning2017},
and \emph{many} more.

A key challenge in online classification is deciding \emph{when} to perform the
classification, \ie after how many packets.
When done too early, the decision basis is too thin to classify a flow confidently.
When done too late, the tardy labeling delays crucial actions. 
For instance, in the context of attacks, late classification might translate to
service disruption (e.g., in the context of Pulse-Wave attacks~\cite{sealsPulseWaveDDoSAttacks2017}).
Moreover, late classification wastes computational resources (e.g., 
tracking and storing features for unnecessarily long). 
Notably, the optimal decision timing varies
across flows. 

\startsquarepar
In an ideal world, network traffic flows would be classified 
whenever the label is sufficiently established -- 
``As Soon As
Possible'' 
\stopsquarepar

\pagebreak
\startsquarepar
\vspace*{0.13cm}\noindent 
(ASAP) -- 
never sooner, never later. 
What makes ASAP classification
hard, though, is that the optimal number of packets to consider for a
classification task heavily depends on the traffic workload, the task
itself, and even the individual flow. In short, a one-size-fits-all solution does not work. 
Nowadays,
in-network classifiers do not support ASAP classification (we discuss related
work at length in~\S\ref{sec:related}). Existing solutions either classify
packets at the end of the flow~\cite{leeSwitchTreeInnetworkComputing2020},
or classify all flows at the same packet 
count~\cite{hullarEarlyIdentificationPeertoPeer2011,bernailleEarlyApplicationIdentification2006,gomezsenaEarlyTrafficClassification2009,bernailleTrafficClassificationFly2006}, 
ignoring
whether classification is possible on a \emph{per-flow} basis.
As our evaluation shows, both approaches are suboptimal in practice.
\stopsquarepar

\startsquarepar
\myitem{\system} In this paper, we describe \system, a system which enables
programmable data planes to perform real-time, ASAP inference, accurately and
at scale, according to supervised machine learning (ML) models. \system takes as
input a labeled dataset (e.g., an annotated traffic trace) and automatically
trains a \PPPP{}-based~\cite{bosshartP4ProgrammingProtocolindependent2014} online 
classifier that can run
directly in the data plane 
and accurately infer labels
on live traffic. Despite being performed in the data plane and
as-soon-as-can-be, \system's inference is accurate. In fact, we show that it is
as accurate as an offline, software-based classification using state-of-the-art
ML frameworks such as \texttt{scikit-learn}~\cite{scikitle17:online}.
\stopsquarepar

\startsquarepar
\myitem{Challenges} Performing accurate ASAP inference in the data plane is
challenging for at least three reasons. First, both the set and values of the 
relevant features change over a flow's lifetime.
As an example, the average packet length may be important to classify young flows, 
but for older flows, the total packet length may be more relevant. The values of
the latter feature diverge as the flow progresses, so the optimal classification rules may change as well. 
ASAP inference therefore requires to dynamically adapt the 
selected features and ML
models over the lifetime of a flow. \system addresses this problem by 
training a \textit{sequence} of random forests---dubbed 
``context-dependent random forests''---that maps to the different phases of a flow. 
The data plane applies this sequence during the life of a flow until the candidate 
label is certain enough; \system decides on a \emph{per-flow} 
basis when the soonest classification is possible.
\stopsquarepar

Second, programmable switches are heavily limited in terms of the operations
they support. In particular, they lack the ability to perform floating point
computations which makes it hard to implement inference procedures for most
ML models (e.g., neural networks) or to keep track of statistical
features (e.g., the average or standard deviation). \system addresses this
problem by: 
\first~classifying traffic according to Random Forest ($RF$) models
whose decision procedures (based on sequential comparisons) fit well within the
pipeline of programmable switches; and 
\second~automatically approximating
statistical features. 
While \system is restricted to $RFs$, 
we stress that $RFs$
are amongst the most powerful and successful ML models currently
available (they can even emulate neural networks~\cite{frosstDistillingNeuralNetwork2017})
and tend to work well in practice~\cite{biauNeuralRandomForests2019,arzaniPrivateEyeScalablePrivacyPreserving2020,arzaniTakingBlameGame2016}. 
Recent works have also
successfully managed to apply $RFs$ to traffic classification tasks (e.g.,
\cite{dukicAdvanceKnowledgeFlow2019,hayesKfingerprintingRobustScalable2016,arzaniPrivateEyeScalablePrivacyPreserving2020,arzaniTakingBlameGame2016}), 
yet without supporting ASAP inference. Finally, $RFs$ are easily
interpretable.

Third, programmable switches have a limited amount of memory (few tens of
megabytes~\cite{jinNetCacheBalancingKeyValue2017}) and do not support dynamic memory management.
Yet, \system needs to compute and store an unknown amount of features during
inference in addition to storing the $RFs$. \system addresses this problem by
considering data-plane constraints while training the $RFs$ and selecting
features that require small amounts of memory. A key insight is that this
optimization does \emph{not} come at the price of classification score. To further deal
with the lack of dynamic memory, \system relies on encoding techniques to pack multiple features in the same register.

We implemented \system in P4$_{16}$ (data-plane inference) and in Python
(training pipeline): given some hardware parameters, \system automatically
synthesizes all P4 code and dynamic CLI input necessary for inference in the
data plane. Our implementation also supports dynamic model reloading.

Our evaluation shows that \system can perform ASAP traffic
classification at line rate, for hundreds of thousands of flows, and with an
classification score that is on-par with software-based solutions. 
We further confirm that all the basic operations required by \system are supported on existing hardware devices (Intel Tofino).

\startsquarepar
Note that \system is a general framework that enables ASAP in-network inference. 
As such, it does \emph{not} remove the need to obtain a representative training dataset. 
As for any ML model, poor input data will result in poor performance. 
We consider the problem of building a representative dataset as orthogonal to this paper.
\stopsquarepar

\myitem{Contributions} To sum up, our main contributions are:
\begin{smallitemize}
	\item An optimization technique for computing random forest models and optimal feature sets tailored to perform as-soon-as-possible classification on programmable data planes (\xref{sec:time_dependent_rfs});
	\item A compilation technique for compiling random forest models to programmable network devices (\xref{sec:compilation_to_p4});
	\item An allocation technique for dynamic memory management available for feature storage (\xref{sec:inference});
	\item A prototype implementation in Python and \PPPP{}, with a confirmation that \system's basic operations run on existing hardware~(\xref{sec:implementation});
	\item \startsquarepar An extensive evaluation using synthetic and real datasets (\xref{sec:evaluation}). \stopsquarepar
\end{smallitemize}

%% file: sec_model.tex
\section{Background} \label{sec:model}\label{sec:background}
In this section, 
we summarize the key concepts of programmable data planes~(\xref{subsec:model_devices}); 
we explain the gist of random
forest classifiers (\xref{subsec:model_randomforest});
and we define the notation (\xref{subsec:model_notation}).

\subsection{Programmable data planes}\label{subsec:model_devices}

We implemented the data plane component of \system in
\PPPP{}~\cite{bosshartP4ProgrammingProtocolindependent2014}, a programming language for network data planes. 
A \PPPP{} program consists of three main building blocks: a
\textit{parser}, which extracts header data from packets arriving at an ingress
port; a \textit{\ma pipeline}, which implements the program's control logic with simple instructions and by applying \ma tables;
and a \textit{deparser}, which assembles and sends the final packet to an
egress port.
We describe key components and limitations of \PPPP{} below.

\myitem{Tables}
Match$\kern 0.7pt$\&$\kern 0.7pt$action tables map keys (\eg packet headers) to actions (\eg set egress port). 
Adding and removing entries to tables is
only possible via the control plane API.

\myitem{Registers}
Registers are stateful objects that are write- and readable both from the
control plane and the data plane. They are organized as arrays of a fixed length
and consist of entries with a fixed width. 
The size of registers needs to be declared at compilation time. 
Since there are no public specifications for the
amount of memory in existing hardware devices, we report the results for units
of 10MB.

\myitem{Operations}
\PPPP{} supports basic operations but no floating point computations or loops.
\system requires the following operations, all of which are supported by
\PPPP{16}~\cite{P416LanguageSpecification2018} and \texttt{bmv2}~\cite{bmv2}: add, subtract,
max, min, bit shift, bit slice.

\myitem{Hardware architecture and resources} Programmable network devices
implement the PISA architecture~\cite{dalyP4Architectures2017}, which contains a 
fixed number
of stages during which \ma tables are applied. The number of these
stages limits the maximum number of tables that can be applied in sequence.

\begin{figure}[ht]
    \renewcommand{\arraystretch}{1}
    $$\begin{array}{lrcl}
        \multicolumn{4}{l}{\textbf{Flows}} \\[0.2em]
        \textit{(Flow)} \qquad			    & \Flow		    & = 	& (\text{src IP},\text{dst IP}, \\
        		    & 		    &                       	& \;\;\text{src port},\text{dst port},\text{prot.}) \\
        \textit{(Packet)}			& \Flow[i]	    &  	& i^{\text{ th}}\text{ packet of the flow} \\
        \textit{(Subflow)}			& \Flow[i:j]    &  	& [\Flow[i], \Flow[i+1], \ldots, \Flow[j-1]] \\
        
        \textit{(Set of flows)}			& \Flowset    &  =	&  \{\Flow_1,  \Flow_2, \Flow_3, \ldots \}\\
        \textit{(Set of subflows)}			& \Flowset[i:j]    & = 	&  \{\Flow_1[i:j],  \Flow_2[i:j], \ldots \}\\
        & 	&  	& \\

       \multicolumn{4}{l}{\textbf{Features}} \\[0.2em]
       \multicolumn{3}{l}{\textit{(Features of }\Flowset\text{)}}		& \features{\Flowset}, \quad \featurematrix \in\mathbb{R}^{m \times n}     \\

       & 	&  	& \\
      \multicolumn{4}{l}{\textbf{Models}} \\[0.2em]
        \multicolumn{3}{l}{\textit{(Random forest)}}			& \RF  \\
        		       
        \multicolumn{3}{l}{\textit{(\RF\xspace trained on }\features{\Flowset[:i]}\text{)}}			& \RF_i  \\
        \multicolumn{3}{l}{\textit{(\RF\xspace applied to }\Flow\text{)}}		& \RF(\Flow)     \\
        \multicolumn{1}{l}{\textit{(Classifier)}}		& \classifier &=& \{\RF_a, \RF_b, \RF_c,\ldots\}, \\ & & & \{a,b,c,\ldots\}\subset\mathbb N     \\

        \multicolumn{3}{l}{\textit{(Decision tree)}}			& DT  \\
        & 	&  	& \\

       \multicolumn{4}{l}{\textbf{Metrics}} \\[0.2em]
       \multicolumn{3}{l}{\textit{(\{True,Predicted\} label)}}			& \rflabel{\Flow}, \rflabel{\RF(\Flow)}  \\
       \multicolumn{3}{l}{\textit{(Certainty)}}			& \certainty{\RF(\Flow)}    \\
       \multicolumn{3}{l}{\textit{($F_1$ score)}}			& \score{\RF(\Flowset)}    \\
       & 	&  	& \\

       \multicolumn{4}{l}{\textbf{Thresholds}} \\[0.2em]
       \textit{(Score threshold)}			& \thrS  & \in & [0,1]  \\
       \textit{(Certainty threshold)}			& \thrC   & \in & [0,1]   \\

    \end{array}$$	
\caption{\system notation and metrics}
\label{fig:notation}
\end{figure}

\pagebreak

\subsection{Random forest classifiers}\label{subsec:model_randomforest}

A random forest ($RF$)~\cite{hoRandomDecisionForests1995} is a supervised ML classifier which
consists of an ensemble of decision trees ($DTs$). To classify a sample, it applies all
$DTs$ on the sample's feature values, obtaining a label 
(\ie estimated class) from each tree. Majority voting results in the final 
label. Additionally, each $DT$ can return a certainty score
$\frac{x_l}{x_{tot}}$ where $x_{tot}$ is the number of training samples that
ended up in the respective leaf node and $x_l$ is the size of the subset of them
with the same label as the current sample. The $RF$'s certainty
for this sample is the mean value of the individual certainties.

\subsection{Notation}\label{subsec:model_notation}

We depict our notation and definitions in~\fref{fig:notation}. We identify a
flow~$\Flow$ as a sequence of packets ($\Flow[i]$ for
$i\in[0,1,\ldots,|\Flow|-1]$) sharing the same 5-tuple (source IP, destination
IP, source port, destination port, protocol). A subflow is denoted as
$\Flow[i:j]$ and consequently the first $n$ packets of a flow are denoted by
$\Flow[0:n]$ or simply $\Flow[:n]$. Sets of flows (as in the datasets that we
use to train the classifier) are denoted as $\Flowset$, sets of subflows as
$\Flowset[i:j]$.

Random forest models are abbreviated by $\RF$, and a model that is trained with
$\Flowset[:n]$ is denoted as $\RF_n$. $\rflabel{\RF(\Flow)}$ denotes the label
that $\RF$ predicts for $\Flow$, $\certainty{\RF(\Flow)}$ the certainty of
the prediction, and $\score{\RF}$ the F$_1$ macro-score of the model.

In training and inference, we use two thresholds: $\thrS$ to denote the minimum
F$_1$ score that is required to accept a model, and $\thrC$ to denote the
minimum certainty required to accept a label.

%% file: sec_overview.tex
\section{\system overview} \label{sec:overview}

\begin{figure}[t]
	\centering\includegraphics[width=0.9\linewidth]{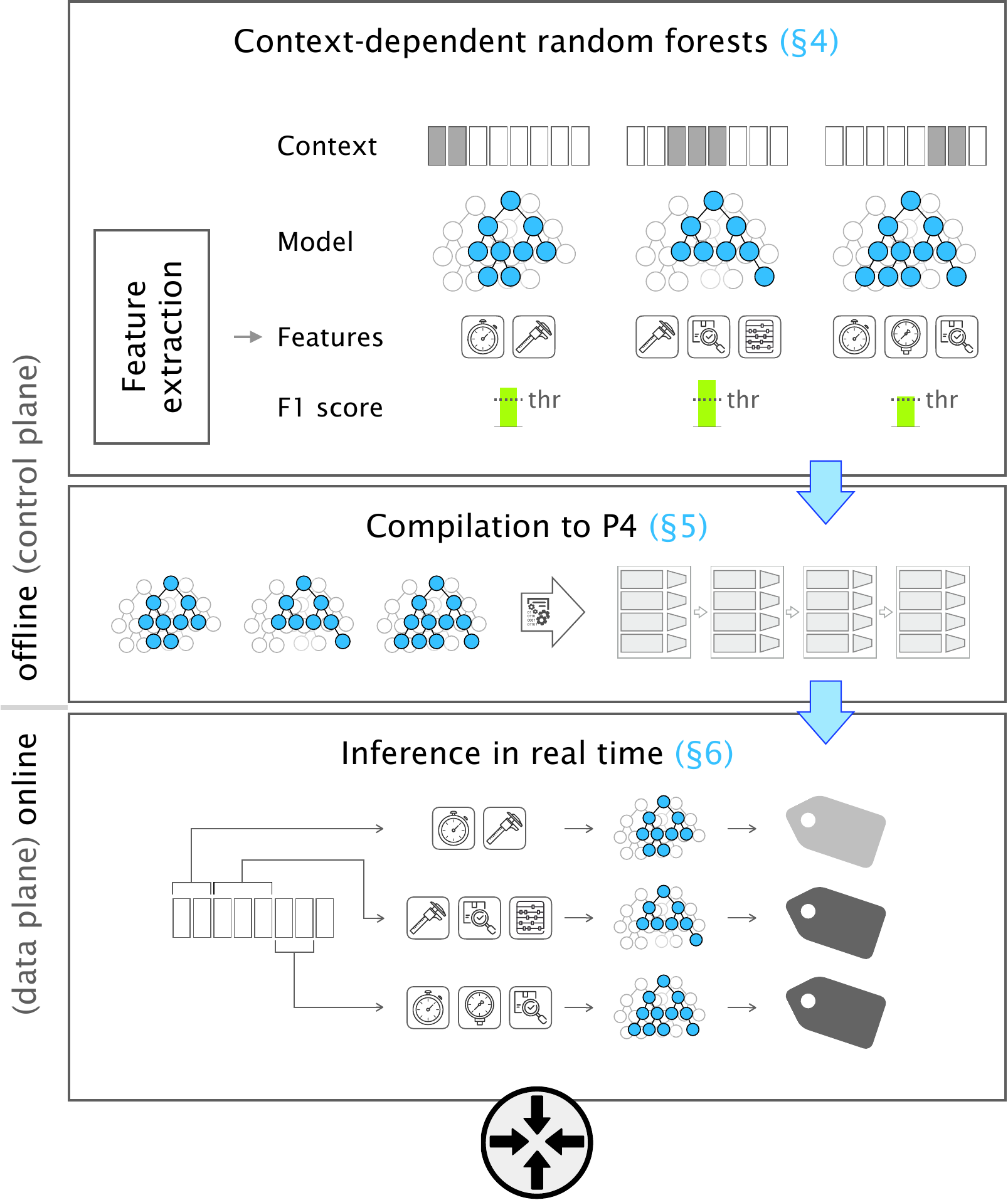}
	\caption{\system overview. From a labeled training dataset, \system
		extracts features and computes context-dependent~$\boldsymbol{\RF s}$.
		Afterwards, it compiles these models to program code and configurations for
		programmable network devices. These programmable network devices then apply
		the models in real time to network traffic.}
	\label{fig:workflow}
\end{figure}

In this section, we describe how \system performs ASAP inference with $\RF$
models entirely in the data plane of a network.

By ASAP inference, we refer to the concept 
of classifiying a flow 
\emph{as soon} as a ``confident'' classification 
is \emph{possible}. 
In the following, we go into more detail on 
\first~\emph{soon} classification and on 
\second~the notion of confidence behind \emph{possible} classification, and then 
\third~put both pieces together into \emph{as-soon-as-possible} inference. 

\myitem{Soon classification}
Observing an additional packet for a given flow can either improve the 
classification confidence or keep it at the same level, since strictly more 
information is given. Naturally, the most certain decision is possible after 
a flow has ended. Therefore, there exists a trade-off between classification 
confidence and classification speed.
\system allows for explicitly selecting a point 
in this trade-off via a threshold on the minimal confidence. 
With this threshold, it can decide on a per-flow level whether it has already 
observed enough packets to make a sufficiently confident decision, or whether it should 
wait for additional packets. 
However, this threshold on confidence is only 
conceptual. In reality, we cannot directly access a single confidence 
metric, but rather measure this concept as described in the next paragraph.

\myitem{Possible classification} In order to assess the confidence for a flow's 
candidate label, \system uses a combination of two thresholds: 
First, a given $\RF$ can report how certain it is about a label for a given sample. 
\system can therefore only accept labels whose certainty exceeds a given threshold
$\tau_c$. 
Second, if the $\RF$ itself has a low quality, it may return a high certainty for 
wrong labels. Thus, \system also applies a threshold $\tau_s$ to the $\RF$ score.

\pagebreak

\myitemNoVspace{ASAP inference} 
Putting the insights on \emph{soon} and \emph{possible} classification together, 
we arrive at ASAP inference. Concretely, it has two ingredients, namely 
\emph{context-dependent $\RF s$} and \emph{certainty-based labeling}.
First, as established, the classification speed may vary 
across flows. Therefore, \system needs to train a sequence of high-quality $\RF s$ that
can classify flows at various stages (\emph{context-dependent $\RF s$}). 
Recall that ``high-quality'' refers to exceeding $\tau_s$.
This is the offline component of ASAP inference. 
Second, at runtime, \system applies the $\RF$ sequence throughout a flow's progress. 
Each $\RF$ reports a candidate label and how certain it is about this label. 
As soon as the certainty exceeds $\tau_c$, \system accepts the label and stops 
the classification process (\emph{certainty-based labeling}). This is the 
online component of ASAP inference. 

\medskip

\startsquarepar
\noindent Overall, this results in the following workflow, also depicted in \fref{fig:workflow}:
Initially, \system takes a
\emph{labeled training dataset} and extracts features which are feasible to
compute in programmable network devices. Based on these features, \system
computes \emph{context-dependent $\RF$s} which exceed a required minimum score,
while minimizing the required hardware memory. Afterwards, \system \emph{compiles}
the $\RF s$ to data-plane programs. Note that we can perform all
these operations offline in the control plane. Finally, \system deploys the
compiled models in programmable network devices which use them to perform
\emph{runtime inference} on the observed network traffic.
\stopsquarepar

\pagebreak
\startsquarepar
\myitemNoVspace{Feature extraction from the training data}
\system uses supervised ML and therefore requires a labeled
dataset to train the $\RF s$ which can support an arbitrary number
of labels (\ie traffic classes). Since the models match subflows of different lengths
($\Flow[:n]$ for length $n\in\mathbb N$),
the training dataset contains
network traffic at packet level (\eg in PCAP format without payload
information).
\stopsquarepar

\system can use up to 18 popular network traffic features per subflow (\ie
$\Flow[:n]$ for subflow length $n$) and it computes them in a way that is feasible
in programmable network devices (\eg replacing averages by moving averages).

\startsquarepar
\myitem{Context-dependent RFs}
During training, \system trains multiple $\RF$s for different contexts,
\ie different numbers of packets that have arrived for a flow. The goal is to 
maximize the score of each $\RF$ while minimizing their required
resources. As these objectives are opposed, \system requires the operator to
provide a score threshold $\tau_s$, turning one objective into a constraint. 
A high $\tau_s$ indicates that
performance is more important than resource optimization (and vice-versa).
\system then reduces the resources allocated to a single $\RF$
until any further reduction would drop its score below $\tau_s$.
\stopsquarepar

\startsquarepar
\system employs several methods to reduce the required resources, \eg memory
space. As an example, the same $\RF$ is shared between multiple contexts if its
score exceeds $\tau_s$ for all of them. Additionally, \system removes redundant
features and reuses them over different contexts to minimize the required
feature memory.
\stopsquarepar

\myitem{Compiling RFs to data plane programs}
After training context-dependent $\RF s$ in software, \system
compiles them for use in programmable network devices. \system implements the
$DTs$ of the $\RF$s as sequences of \ma tables,
where each table contains the nodes of one stage.
This allows us to leverage the pipeline architecture of programmable network
devices and to apply all $DTs$ in parallel. As a result, the entire
model is encoded in memory that can be changed at runtime (\ie table entries),
thus \system allows for a seamless deployment of new or updated models.

Furthermore, \system maximizes memory efficiency by dynamically allocating
memory for required features, circumventing the fact that programmable network
devices do not support dynamic memory management per se. 
In order to overcome this limitation, 
\system applies a
sophisticated encoding technique to store different features in a single
register. On a high level, it reduces the precision
(\ie bit-width) of feature values, concatenates all features to one bitstring,
and stores the feature positions in a register.

\myitem{Runtime inference in the data plane}
At runtime, there are again two opposing objectives. The goal is to classify flows
as early as possible, while maximizing the classification score. In practice,
there is no ground truth available, and thus the prediction score can not be
computed at runtime. Instead,  we take the prediction \emph{certainty}
as a proxy. It is available at the same time as the prediction.

\startsquarepar
Similar to before, these objectives are opposed to each other: Waiting for
additional packets decreases classification speed, yet can increase the
classification certainty, as more information becomes available. Thus, \system
requires the operator to provide a certainty threshold $\tau_c$, again turning
one objective into a constraint. 
A subflow is
only classified if the classification certainty is above 
\stopsquarepar

\pagebreak
\noindent
$\tau_c$, otherwise
\system waits for additional packets. As long as the
certainty score is below a
given value, \system continues to track the flow, storing intermediate results
such as averages in memory. As soon as the certainty exceeds $\tau_c$, the
classification label is accepted, memory freed, and the flow is no longer
tracked.

For example, assume that \system has trained three $\RF$s and compiled 
them to the data plane: 
$\RF_3,\ \RF_5,\ \RF_{10}$, 
trained on $\Flowset[:3],\ \Flowset[:5],\ \Flowset[:10]$ 
respectively, and the operator has chosen $\tau_c = 0.9$.
Now, packets of a single flow arrive.
At packet 3, $\RF_3$ computes a candidate label for the flow, 
and returns a certainty of $0.8$ for this label. Hence, the label is not accepted
and \system keeps tracking the flow. 
At packet 5, $\RF_5$ computes the next candidate label with a 
certainty of $0.91$, thus exceeding $\tau_c$. 
Therefore, \system accepts the label. 
This highlights the trade-off between speed and certainty: 
Had $\tau_c$ been $0.75$, the label would already have been accepted 
at packet 3.

%% file: sec_time_dependent_random_forests.tex
\section{Context-dependent random forests} \label{sec:time_dependent_rfs}

In this section, we describe how we build the
first component for ASAP inference, namely a classifier based on $\RF$
models. On a high level, we do so with the following optimization problem: The classifier should
\textit{maximize} speed, \ie classify a flow after few packets (\emph{as soon \ldots});
and it should \textit{provide} a classification score above a given threshold (\emph{\ldots as possible}).

Moreover, it should \textit{minimize} memory usage,
and use models that \textit{can run} on programmable network devices. In
\xref{subsec:rf_optimization}, we describe the optimization problem in detail;
in \xref{subsec:rf_greedy}, we explain how \system approximates the optimal
solution; and in \xref{subsec:rf_training}, we describe how \system trains
context-dependent $\RF s$.

\subsection{Optimization problem}\label{subsec:rf_optimization}

\startsquarepar
\system computes and applies classifiers according to the following optimization
problem. \textit{Given a labeled dataset $\Flowset$ and a minimal threshold score
    $\thrS$, find a classifier $\classifier$ such that
    $\score{\classifier}\geq\thrS$ and it is feasible to run $\classifier$ in
    programmable network devices while minimizing the required memory and maximizing
    the classification speed.}
\stopsquarepar

\myitem{Objective I: Minimizing memory usage} As programmable network
devices dispose of very limited memory resources, \system minimizes the amount of
per-flow memory. This size directly relates
to the number of concurrent flows that \system can classify.

\myitem{Objective II: Maximizing classification speed} For many applications,
online traffic classification is only useful if an ongoing flow is classified
within its first few packets. Therefore, \system classifies flows as early as
possible.

\myitem{Constraint I: Guaranteed score} \system produces a classifier that
exceeds a given score threshold, in terms of the F$_1$ macro score
(\ie the unweighted average over the
F$_1$ scores of each class~\cite{sklearnm61:online}).

\myitem{Constraint II: Feasibility in hardware} \system is designed to work in
hardware devices supporting the \PPPP{} language and with realistic
specifications (\cf \xref{subsec:model_devices}).

\newpage
\startsquarepar
\myitemNoVspace{Optimal solution}
Maximizing classification speed implies that a flow needs to be classified after
a
few packets and without knowledge of the packets that will arrive afterwards.
However, the packets that arrive afterwards can have an impact on the feature
values and make individual features more or less relevant. Thus, waiting
for one more packet (\ie reducing the classification speed) could allow for using
fewer features (\ie increasing the memory efficiency).
\stopsquarepar

Furthermore, the time at which a flow can be classified can differ for each
flow, even if they belong to the same class.
Therefore, finding the optimal solution would require to cover a search space of
$\bigO{\textit{\# flows}\times\textit{flow length}\times\textit{\# features}}$.
Because this is infeasible, \system approximates the optimal solution through a
greedy algorithm which we describe below.

\vspace{-0.1cm}
\subsection{\system greedy algorithm}\label{subsec:rf_greedy}

\startsquarepar 
Instead of searching a globally optimal classifier, \system generates multiple
-- locally optimal -- $\RF$ models and combines them. This reduces the
size of the search space from exponential to linear (\ie finding the best model
in each context). A context of a flow is defined as the first $x$ packets of a
flow (\ie $\Flow[:x]$). For each context, \system therefore solves the following
variant of the optimization problem: \textit{Given a labeled dataset
    $\Flowset[:x]$ and a minimal threshold score $\thrS$, find a classifier
    $\RF$ such that $\score{\RF(\Flowset[:x])}\geq\thrS$ and it is feasible to
    run $\RF$ in programmable network devices while minimizing the required memory
    and maximizing the classification speed.}
\stopsquarepar 

\myitem{Guarantees} Each of the context-dependent $\RF s$ is
locally optimal in that it has a classification score
$\score{\RF(\Flowset[:x])}\geq\thrS$, but only uses the set of 
features necessary to exceed $\thrS$, thus minimizing the memory consumption
of the per-flow feature values. 
We do not directly optimize the combined classifier, but for the special case
where the label of the first $\RF$ is accepted in any case (\ie irrespective of
the certainty), the overall score is equal to the score of the first
model and therefore $\geq\thrS$. By combining multiple models, the overall
score exceeds $\thrS$, as we show in the evaluation.

\vspace{-0.1cm}
\subsection{Training context-dependent RFs}\label{subsec:rf_training}

\startsquarepar
\system trains context-dependent $\RF s$ in the following steps:
\first it extracts the features;
\second it groups redundant features;
\third it selects the optimal representative feature from each group;
\fourth it searches for the optimal model for a given set of features
$\features{\Flowset[:x]}$, and increases $x$ until it finds
a sufficiently good $\RF_x$;
\fifth it retrains $\RF_x$ with the selected features and adds it to the final classifier $\classifier$; and
\sixth it tries to reuse $\RF_x$ on $\Flowset[:y]$, for increasing $y > x$,
until the score drops below $\thrS$.
If the score has dropped below $\thrS$ at $\Flowset[:y]$,
\system checks whether one of the previously extracted models can be used. If so, it adds the best of them
($=: \RF_k$)
to $\classifier$ and jumps
to \sixth.
If the score of $\RF_k$ was below $\thrS$, it jumps to \third
instead.
\stopsquarepar

\begin{figure}[t]
    \centering\includegraphics[width=0.95\linewidth]{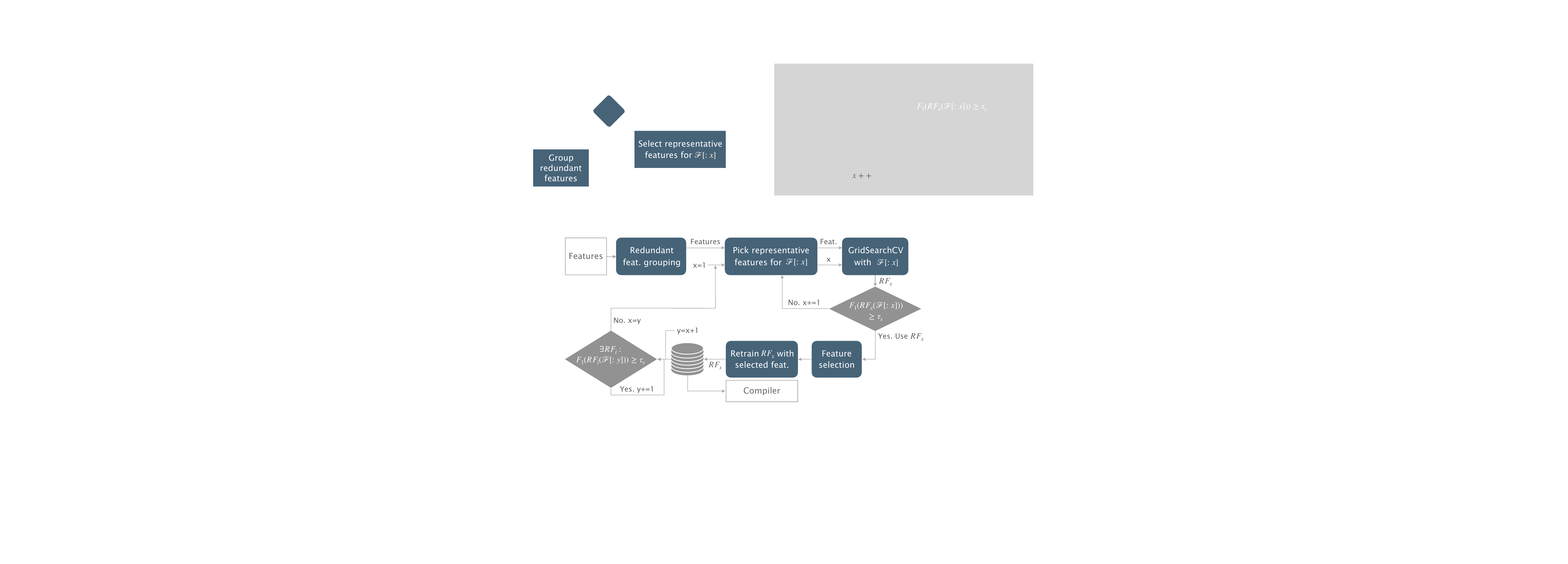}
    \caption{\system training algorithm} 
    \label{fig:algorithm_flowchart}
\end{figure}

\startsquarepar
The following paragraphs describe each step 
(cf. also \fref{fig:algorithm_flowchart}). 
\stopsquarepar

\myitem{Network traffic features}
\system's feature extraction component extracts 18 popular
network traffic features inspired by CICFlowMeter~\cite{lashkariAhlashkariCICFlowMeter2022} and listed in
\tref{tab:supported_features}. In contrast to~\cite{lashkariAhlashkariCICFlowMeter2022}, \system
extracts features per subflow (\ie $\Flow[:n]$ for $n=1,2,3,\ldots$) in order
to allow context-dependent models based on subflows, and it respects the
limitations of programmable network devices.
In this stage, \system also splits the features into a train and test
dataset with 9:1 proportions (also for each individual label sort). 
\newpage

\startsquarepar
\myitemNoVspace{Grouping redundant features}
It could happen that the $\RF$ training selects a feature that is much harder to 
compute than an alternative, directly correlated feature. 
In order to avoid such situations, \system clusters features into groups
that carry very similar information, from which it later picks one representative (\cf next step). 
Concretely, it computes the mutual
information $\operatorname{I}$ among the training features (\eg feature $f_i$ and feature
$f_j$) and constructs the normalized distance metric $ d(f_i, f_j) = 1 -
    \frac{\operatorname{I}(f_i; f_j)}{\operatorname{H}(f_i, f_j)}$ with
$\operatorname{H}(X, Y)$ being the joint entropy of discrete random variables
$X$ and $Y$. This results in a distance matrix $D$ with entries $d_{ij} = d(f_i,
f_j)$. It runs the DBSCAN
algorithm~\cite{esterDensityBasedAlgorithmDiscovering1996} on this matrix to 
cluster the features. 
\stopsquarepar

\myitem{Selecting representative features}
Given the groups of redundant features, \system selects a
representative feature for each group. Since the features within a group carry
similar information, we can use additional trade-off metrics for the selection
while expecting a similar classification score. We leverage this to optimize for
memory usage, convergence speed, and number of distinct features:

\begin{itemize}
    \item \textit{Memory usage:} \system prefers features which require less
          memory. To incorporate this, we define a metric $m_{\text{m}} = \text{(feature
                  size in bits)}$. We derive the feature size from \texttt{bmv2}~\cite{bmv2} (for
          averages, we add 2 bits to the field size for more accuracy; for counters we
          assume a maximum size of 127, i.e. 7 bits; stateless features do not require
          memory).
    \item \textit{Convergence speed:} \system prefers features whose values
          converge within few measurements (packets). The value of this metric
          ($m_{\text{c}}$) is determined by the number of packets that are needed in order
          to compute it (\eg $m_{\text{c}}=1$ for the Pkt Len, $m_{\text{c}}=2$ for
          IAT and $m_{\text{c}}=3$ for the average IAT).
    \item \textit{Number of distinct features:} \system prefers features, which
          are used in previous models in order to reduce the number of features it needs
          to compute and store: \startsquarepar $m_{\text{d}} = 0 \text{ if a previous model used
                  this feature, else }m_{\text{d}}=1$ \stopsquarepar
\end{itemize}

\system computes these metrics for each feature in a group; normalizes them;
computes a weighted average ($w_m \cdot m_{\text{m}} + w_c \cdot m_{\text{c}} +
    w_d \cdot m_{\text{d}}$); and selects the feature with the lowest total score as
the representative of the group. The weights are initially set to $1$ ($w_m$ and
$w_c$) and $0.5$ ($w_d$) and then decrease linearly in the number of models
towards $0$. This mimics how the importance of metrics changes over time: For
an $\RF$ that appears later in the $\RF$ sequence, it becomes more important that the same features are reused in
comparison to convergence speed and memory usage.

    \begin{table}[!t]
        \footnotesize
        \renewcommand{\arraystretch}{1.2}
        \begin{center}
            \begin{tabular}{@{}ll@{}}
                \toprule
                \multicolumn{2}{@{}l}{\textit{Stateful features}}                        \\
                \hspace{.2cm}IAT        & Packet inter-arrival time (min, max, avg)      \\
                \hspace{.2cm}Pkt Len    & Packet length (min, max, avg, total)           \\
                \hspace{.2cm}Pkt Count  & Number of packets                              \\
                \hspace{.2cm}Flag Count & TCP flag counts (SYN, ACK, PSH, FIN, RST, ECE) \\
                \hspace{.2cm}Duration   & Time since first packet                        \\
    
                \multicolumn{2}{@{}l}{\textit{Stateless features}}                       \\
                \hspace{.2cm}Port       & TCP/UDP port (source, destination)             \\
                \hspace{.2cm}Pkt Len    & Length of current packet                       \\
    
                \bottomrule
            \end{tabular}
            \caption{Supported features}
            \label{tab:supported_features}
        \end{center}
    \vspace{-0.7cm}    
    \end{table}

\pagebreak

\myitemNoVspace{Model search}
\system optimizes the structure of the $\RF s$ on the training dataset,
and across three parameters:
\first the maximum depth of the trees; \second the number of trees; and \third
the weights of the classes during training. \first and \second can be defined
such that the model fits onto a particular programmable network device. \third
allows \system to handle imbalanced datasets. 
\system
optimizes the F$_1$ macro score through a 6-fold cross validation, where the
folds are chosen such that the classes are represented with the same percentage
across all folds. Since it uses the F$_1$ macro score, the individual 
F$_1$ scores for all classes
are weighed equally irrespective of their size, which further addresses
the issue of imbalanced datasets.

It then constructs a model with the parameter choice that proved best, 
and retrains the model on the entire training dataset. It then checks
whether the score on the test dataset exceeds $\thrS$; if so, it selects the
model into the sequence. 

\startsquarepar
\myitem{Model optimization}
Once \system has found a model with a score $\geq\thrS$, it selects the minimal
number of features necessary to achieve $\thrS$.
To do this, \system ranks the features according to the mean decrease
in impurity (MDI)~\cite{louppeUnderstandingRandomForests2015}.
It first trains the model with the most important feature, then with
the two most important features and so on until the score of the trained model
is $\geq\thrS$.
This type of memory optimization is a local optimization for the \emph{current}
$\RF$. An optimization across the sequence of all $\RF$s, \ie $\classifier$,
would require a simulation of how each feature selection fares in the future.
\stopsquarepar

\myitem{Longest-possible model reapplication}
\system reapplies the most recent $\RF$ ($ =: \RF_i$) to $\Flowset[:j]$, for
increasing $j > i$, until the score drops below $\thrS$. If the score has
dropped below $\thrS$ at $\Flowset[:j]$, \system tests all previously extracted
$\RF$s. If the best of them ($=: \RF_k$) has a score above $\thrS$, it reuses
$\RF_k$ ($=: \RF_j$), and appends $\RF_k$ to the extracted sequence $\classifier$.
It then reapplies this $\RF_k$ for as long as possible as well. If the score of
$\RF_k$ was below $\thrS$, \system again starts the search for a new model.

\begin{table}[!b]
    \footnotesize
    \renewcommand{\arraystretch}{1.2}
    \begin{center}
    \begin{tabular}{@{}lll@{}}
    \toprule
    \textbf{Property} & \textbf{Output type} \\ 
    \midrule
	\textit{Features} & & \\
    \hspace{.2cm} Extraction and computation & Code\\ 
    \hspace{.2cm} Memory assignment & Configuration\\ 
	\textit{Flows} & & \\
    \hspace{.2cm} Feature memory per flow & Code\\ 
    \hspace{.2cm} Number of trackable flows & Code\\ 
	\textit{Random forests} & & \\
    \hspace{.2cm} Maximum dimensions & Code\\ 
    \hspace{.2cm} Models & Configuration\\ 
    \hspace{.2cm} Classification thresholds & Configuration\\ 
    \bottomrule
    \end{tabular}
    \caption{Compiler output} 
    \label{tab:compiler_output}
    \end{center}
\end{table}
\vspace{-1cm}

%% file: sec_compilation_to_p4.tex
\newpage
\section{Compiling random forests to the data plane}\label{sec:compilation_to_p4}

This section explains how \system compiles context-dependent $\RF$
models to code and configuration for programmable network devices. We give
an overview over the outputs of the \system compiler
(\xref{subsec:compiler_output}), describe how \system encodes
context-dependent $\RF s$ in \PPPP{}
(\xref{subsec:compiler_randomforest}), and how it optimizes memory allocation
(\xref{subsec:compiler_memory}).
\medskip

\subsection{Compiler output}\label{subsec:compiler_output}

\system compiles context-dependent $\RF s$ to two types of output:
\textit{program code}, which runs on \PPPP{}-programmable network devices, and
\textit{program configuration} which specifies the program's behavior.

The key difference between code and configuration is that changing the code
requires a restart of the device while changing the configuration can happen
on-the-fly. \system compiles $\RF s$ to configuration so that
they can be updated at any time. It only encodes those parts in code which
\PPPP{} does not allow to be configured at runtime (\eg the total size of 
feature memory). Configuring \PPPP{} applications is possible in two main ways: through entries
in \ma tables and through values in stateful memory (\ie registers). 

Table~\ref{tab:compiler_output} summarizes the outputs of the compiler and
specifies whether they are in form of program code or configuration. 
The following sections describe each output in more detail.

\subsection{Random forests in \ma tables}\label{subsec:compiler_randomforest}

$\RF s$ consist of multiple $DTs$ which output a label and a
certainty for each given sample. An $\RF$ model then computes the final
label through majority voting by all the $DTs$ (\cf
\xref{subsec:model_randomforest}).

\myitem{Encoding DTs}
Applying $DTs$ resembles
the \ma pipeline in programmable network devices (\ie the PISA
architecture~\cite{dalyP4Architectures2017}). In the latter, a packet 
traverses several stages of the
pipeline. Each stage can read and modify attributes and influence the
processing in the following stage. Applied to $DTs$, each stage
represents one level of the tree and the packet conveys the feature values. 

In each level of the $DT$, the model compares one of the features
against a threshold specified by the matching node. This translates directly to
\ma pipelines: tables match on the ID of the current node, define the
threshold as well as the feature to compare, and -- depending on the result of
the comparison -- the ID of the next node. At the end, leaf nodes assign the
label and its certainty.

\begin{figure}[t]
	\centering\includegraphics[width=0.8\linewidth]{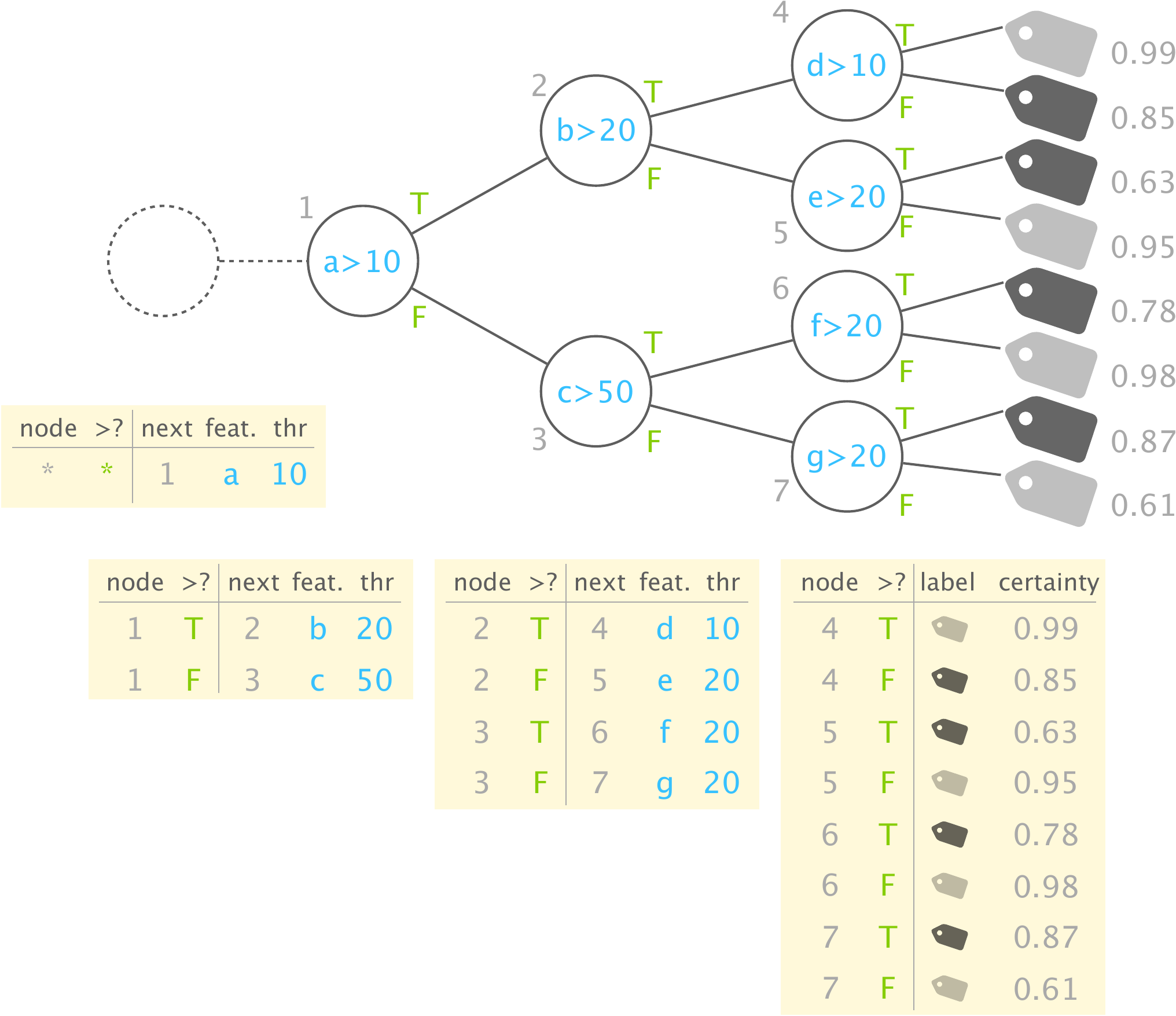}
	\caption{Encoding $\boldsymbol{\RF s}$ in \ma tables}
	\label{fig:encoding_random_forest}
\end{figure}

As illustrated in \fref{fig:encoding_random_forest}, \system encodes decision
trees in table entries of the following form:
\color[HTML]{2d5403}
\vspace{-.2cm}
\begin{multline*}
(\texttt{prev node}, \texttt{ prev comparison result}) \mapsto \\
(\texttt{next node}, \texttt{ feature to compare}, \texttt{ threshold})
\end{multline*}
\color{black}
where\xspace\xspace
\color[HTML]{2d5403}\texttt{prev comparison result}\color{black}\xspace
\xspace is \xspace
\color[HTML]{2d5403}\texttt{True}\color{black}\xspace
\xspace iff the feature was
larger than the threshold of the previous node.
Leaf nodes map features to a label and a certainty:  \color[HTML]{2d5403}
\vspace{-.2cm}
\begin{multline*}
(\texttt{prev node}, \texttt{ prev comparison result}) \mapsto \\
(\texttt{label}, \texttt{ certainty})
\end{multline*}
\startsquarepar \color{black}
\myitemNoVspace{Encoding RFs}
Developing this approach further from one to multiple decision
trees in an $\RF$ is straightforward: \system encodes each $DT$
in its own tables (\ie one table per $DT$ and level). This allows to
apply all $DTs$ in parallel. \system then combines the labels
and certainties of all individual $DTs$ to a final label and its certainty.
\stopsquarepar

\pagebreak \color{black}
\myitemNoVspace{Encoding context-dependent RFs}
Encoding multiple context-dependent $\RF s$ is analogous to encoding one
$\RF$ as described above. It does not require additional tables because
only one $\RF$ is applied to each packet.
\system uses a table to map the current flow's packet count to the 
applicable $\RF$ for that phase. 

\subsection{Allocating feature memory}\label{subsec:compiler_memory}

\startsquarepar
\system applies two techniques in order to dynamically allocate the optimal
number of bits for each feature despite the fact that \PPPP{} does
not support dynamic memory management: \first it adjusts the precision with
which it stores each feature depending on the $\RF s$; and \second
it concatenates all required features into one bitstring. 
\stopsquarepar

\myitem{Allocating the optimal number of bits per feature}
$\RF s$ do not require the absolute value of a feature. Instead,
they only depend on the result of comparing feature values with thresholds.
\system leverages this for saving memory by reducing the precision and the range
of the stored features in a way that allows precise comparisons with the
thresholds in all models.

For positive feature values with strictly positive thresholds, and a 
given minimum comparison accuracy $a$, the minimum needed amount 
of bits $b$ is 
\begin{equation}
b = \left \lfloor{\log_2{\frac{2 \cdot t_{max}}{t_{min} \cdot 0.5 \cdot a}}}\right \rfloor + 1
\end{equation}

where $t_{max}$ and $t_{min}$ are the maximum and minimum thresholds with which the
feature is 
compared\footnote{Generally, one needs $\lfloor{\log_2{v}}\rfloor + 1$ bits to 
encode a value $v \in \mathbb{N}$. For a successful comparison with $t_{max}$, 
we must provide one bit more than used by $t_{max}$.
Thus, the maximum encoded value is $2 \cdot t_{max}$, in \textit{units of} the smallest stored value: 
analogously, $\frac{1}{2} \cdot t_{min}$, further
decreased with the given precision parameter $a$.}. 
Accordingly, the amount of bits by which the feature in P4 needs to be shifted is

\begin{equation}
s = \left \lfloor{\log_2{t_{min} \cdot 0.5 \cdot a}}\right \rfloor
\end{equation}

For counter features it holds that $a = 1$ and $t_{min}=1$ because they are integers 
and need to be able to count from $1$.
If several $\RF s$ use the same features, the maximum and minimum
thresholds are computed over all of them.

\emph{Example:} If $\RF_3$ and $\RF_5$ (and no other $\RF s$ in $\classifier$)
use the average of the packet length as a feature, \system looks for the overall
maximum and minimum threshold $t_{max}$ and $t_{min}$ that could be applied to
this feature in $\RF_3$ and $\RF_5$, as an example $1234.5$ and $67.8$. If the
comparison accuracy is $0.01$, the resulting number of bits is $\left
\lfloor{\log_2{\frac{2 \cdot 1234.5} {67.8 \cdot 0.5 \cdot 0.01}}}\right \rfloor
+ 1 = 13$.

\myitem{Encoding all features in one bitstring}
As explained above, \system computes the number of bits that are required to
store each feature. Instead of saving every feature in a separate register,
\system concatenates all values to one bitstring, which allows for dynamically
allocating the per-flow memory and changin $\RF s$ without
re-compiling the program. In order to specify the location of each feature
in this bitstring, \system stores the positions (start and end index) in a
register on the programmable network device.

%% file: sec_inference.tex
\section{ASAP Inference in the data plane}\label{sec:inference}

\startsquarepar
We now describe \system's data plane component, 
the runtime aspect of ASAP inference. Its code is
partially automatically generated by the compiler (\cf
\xref{sec:compilation_to_p4}). After an overview over the entire pipeline that a
packet passes in a \system device, we give special attention to how \system extracts the features
in the data plane, how it manages memory for them, and how it performs
certainty-based labeling. 
\stopsquarepar

\subsection{Pipeline overview}

\begin{figure}[b]
	\captionsetup{belowskip=0pt}
	\centering\includegraphics[width=0.9\linewidth]{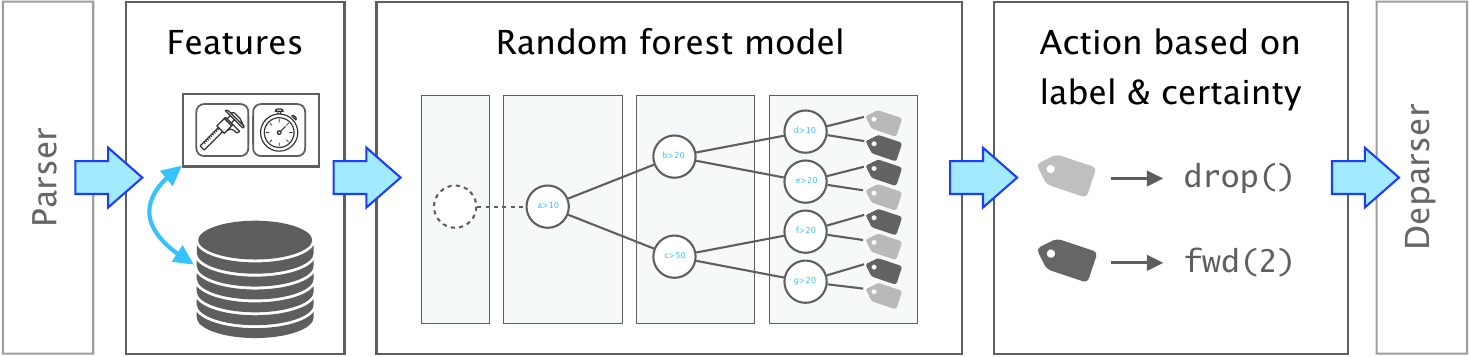}
	\caption{Packet processing pipeline in the data plane}
	\label{fig:p4_pipeline}
\end{figure}

\fref{fig:p4_pipeline} illustrates the packet processing pipeline which \system
runs in programmable network devices. Here, we summarize each processing step.
The ensuing subsections provide detailed information.

\myitem{Parsing}
\system extracts the Internet and transport layer header, as it contains
information for some of the features. Further, it uses the 5-tuple (source IP,
destination IP, source port, destination port, protocol) as an identifier for a
flow.

\startsquarepar
\myitem{Feature lookup}
\system computes several hashes of the flow ID and uses them to derive
the storage location for the flow (\S\ref{subsubsec:flow_mem_mgmt}).
\stopsquarepar

\myitem{Updating features}
\system computes the updated features based on the received packet (\S\ref{subsubsec:feature_computation})
and stores
them in registers. The features are maintained in a compressed form (\S\ref{subsec:feature_mem_mgt}).

\myitem{Applying the RF model}
\system sends the packet through the series of \ma tables which encode
the $DTs$ of the $\RF$~(\cf~\xref{sec:compilation_to_p4}). The
packet count (one of the previously extracted features) determines the model
used for the classification.

\myitem{Deriving the label and certainty}
\system aggregates the labels (by majority-voting) and certainty scores 
(by summing\footnote{Since the number of trees per $\RF$ is known, we can compare 
the sum with the non-normalized certainty threshold and thus avoid computing the average with division.}) 
from the individual
$DTs$ in the current $\RF$ to one label and certainty.

\pagebreak
\myitemNoVspace{Certainty-based labeling} \system uses $\tau_c$ to determine whether it trusts 
the label or not, \ie
whether a definitive classification of this flow is already possible or not (\S\ref{subsec:certainty_based_labeling}). 

\myitem{Further actions based on label and certainty} 
\system can apply arbitrary additional actions based on the label and the certainty, 
since this information is stored in the metadata of each packet. 
For example,
traffic labeled as malicious could be sent to a security device for further
inspection, traffic labeled as VoIP could be forwarded with higher priority, 
or flow IDs labeled as file transfers could be sent to a monitoring system. 
Adding such actions is straightforward and out of scope of this paper.

\subsection{Feature extraction} \label{subsubsec:feature_computation}

\system supports different types of features (\cf~\tref{tab:supported_features}): minimum
values, maximum values, average values, counters, sums, differences and
stateless metadata of the current packet. Most of them are straightforward to
implement in \PPPP{}. 

However, the lack of division and floating point
operations makes it challenging to compute average values. 
Because of this, \system uses the exponentially weighted moving average
(EWMA) as an approximation. The EWMA computes to 
\vspace{-0.08cm}
$${\displaystyle S_{t}={\begin{cases}Y_{1},&t=1\\\alpha \cdot Y_{t}+(1-\alpha
)\cdot S_{t-1},&t>1\end{cases}}}$$
where $S_t$ is the updated EWMA,  $S_{t-1}$ the previous value, and $Y_t$
are the values to be averaged. The constant $\alpha \in [0, 1]$ determines how
much the current value influences the average, \ie how fast past values
are discounted. Since multiplications are not possible in many \PPPP{}
targets, we use $\alpha =0.5$ such that multiplications can be replaced by bit shifts.

\subsection{Feature memory management}\label{subsec:feature_memory}

We now describe how \system solves the two challenges of
dynamically and efficiently assigning
\first~flows to a fixed number of memory cells, and
\second~parts of these memory cells to features.

\startsquarepar
\system manages memory in two dimensions: \textit{per-flow memory} is a
bitstring of fixed size associated with each flow that \system currently
classifies. This bitstring contains \emph{per-feature memory} blocks of dynamic
size. Each of these blocks contains one stateful feature and \system chooses the
size of the block depending on the precision which the model requires for the
respective feature (\cf \xref{subsec:compiler_memory}). 
\stopsquarepar

\subsubsection{Per-flow memory management}\label{subsubsec:flow_mem_mgmt}
At compile time, \system creates an array of registers to later store per-flow
features. Each entry in this register contains the \textit{flow ID} (a 32bit hash of
the flow's 5-tuple)\footnote{Experiments with the CAIDA traces~\cite{TheCAIDA64:online} show that the probability of a hash collision is only $\approx 3.085 \cdot 10^{-5}$}, the \textit{timestamp} of the flow's last packet, the
\textit{packet count} saying how many packets the switch received for this flow,
and dynamically split space for multiple \textit{feature values}.

\myitem{Efficient allocation of flows to memory cells}
Typically, the number of registers is much smaller than the number of 
possible flows (5-tuples). Since there are no linked lists
or the like in \PPPP{}, \system implements an allocation strategy using hash-based
indices.

\pagebreak
Computing the index of a flow by using only one hash function is not efficient
because of hash collisions (\ie different flows hash to the same index). To
avoid this, \system computes multiple hashes of the flow ID using different hash
functions. It then checks the register array at these indices for \First whether
it contains this flow ID; and \Second for whether its rows at these indices are
usable. A row is usable if the slot is empty, or the last packet of the flow in
the slot was more than a predefined timeout ago. If \First fails, \system uses
the first-best usable row according to \Second to store the flow ID. If both
conditions \First and \Second fail, \system cannot store stateful features and
forwards the packet without classification. However, it adds a flag, which
allows other devices to determine whether a flow was classified or not. That
flag is stored in the Reserved Bit in the IP header. \system implements this as
a naive way of distributing the classification. More sophisticated approaches
are out of scope for this paper, but we discuss ideas in \xref{sec:discussion}.

\subsubsection{Per-feature memory management}\label{subsec:feature_mem_mgt}

Besides the attributes that \system needs to store for all random
forest models (flow ID, last timestamp and packet count), \system splits the
remainder of the per-flow memory dynamically into fields in which it stores the
other features. As explained in \xref{subsec:compiler_memory}, this allows
\system to deploy a new $\RF$ model based on other features without
re-compiling the \PPPP{} program and without interrupting the network. 

\startsquarepar
Concatenating all features into one bitstring and extracting them from the
bitstring is possible using the bit slice operator in \PPPP{16}, or 
bit shifts in \PPPP{14}. As described in \xref{subsec:compiler_memory}, the
compiler defines the location of each feature in the bitstring and stores it in
registers.
\stopsquarepar

\subsection{Certainty-based labeling}\label{subsec:certainty_based_labeling}

As described in \xref{subsec:compiler_randomforest}, \system implements
$\RF s$ through a series of \ma tables. They specify the
feature and the threshold to compare in each node, as well as the following node
depending on the result of the comparison.

\startsquarepar
Each arriving packet triggers a classification attempt of the respective flow,
with the currently applicable $\RF$ (if there is one, which might not be
the case for the first few packets). \system trusts the classification attempt if
the certainty for the predicted label is above a threshold ($\thrC$).
Upon a trusted classification, \system empties the space allocated to that 
flow such that it can store another flow.
\stopsquarepar

%% file: sec_implementation.tex
\section{Implementation}\label{sec:implementation}

\startsquarepar
In this section, we outline the implementation of the training component, the
data plane program, and the hardware prototype.
\stopsquarepar

\myitem{Feature extraction and training}\label{subsec:implementation_training}
Our implementation for extracting features, training context-dependent $\RF s$, 
and compiling them to \PPPP{} consists of about 17~000 lines
of Python code. We use \texttt{dpkt}~\cite{dpktonline} for parsing network
traffic and extracting features and 
\texttt{scikit-learn}~\cite{scikitle17:online} for computing $\RF$
models.

\myitem{Inference in the data plane}\label{subsec:implementation_p4}
The data plane component consists of about 1500 lines of \PPPP{16} code
(for an $\RF$ of 32 trees with maximum depth 10). This code runs in the
behavioral model~\cite{bmv2}, and we tested it using Mininet~\cite{lantzNetworkLaptopRapid2010}\footnote{Currently, the bmv2 implementation does not support classification for offline flows, \ie after timeout.}.

\pagebreak

\startsquarepar
\myitem{Hardware prototype}\label{subsec:implementation_tofino}
In addition to an implementation in \PPPP{16}, we verified that the basic operations of \system run in real hardware. To do so, we implemented an $\RF$ on Intel Tofino \cite{tofino_web}. Our implementation supports features of type counter (\eg for ACK counter) and max/min (\eg
packet size) and $\RF s$ up to depth 4.
\stopsquarepar

%% file: sec_evaluation.tex
\section{Evaluation} \label{sec:evaluation}

In this section, we use real and synthetic datasets
(\xref{subsec:evaluation_datasets}) to showcase the functionality
of \system. 
We first confirm that the mechanisms behind ASAP inference work 
as intended: we show that 
\first~\system's training algorithm
successfully creates context-dependent $\RF s$,
the offline
component of ASAP inference (\xref{subsec:evaluation_synthetic});
and we establish that
\second~\system successfully classifies each flow as-soon-as-possible using the
certainty threshold, the online component of ASAP inference~(\xref{subsec:evaluation_speed}). 
We also find that \system achieves a high score
both in software and hardware thanks to an accurate translation
(\xref{subsec:evaluation_accuracy}), 
while it uses  little memory~(\xref{subsec:evaluation_scalability}).

\subsection{Datasets and methodology}\label{subsec:evaluation_datasets}

\startsquarepar
Below, we describe each used dataset, the applied preprocessing steps, 
the settings for the machine learning models, and 
the simulation technique. 
We use three datasets (summarized in Table~\ref{tab:evaluation_datasets}): 
a synthetic dataset that we generated ourselves and two public ones.
\stopsquarepar

\myitem{Synthetic dataset} The synthetic dataset consists of
artificial feature
values for 100'000 flows, each with nine packets. We directly generate
the features for subflows  $\features{\Flowset[:i]}$, $i \in {1, \ldots, 9}$, and the corresponding
labels with \texttt{sklearn}~\cite{sklearnd30:online}, such that different
features are relevant in different phases of the flows. At packet count 4, 
we simulate that features $\feati{1}$ - $\feati{3}$ are redundant to the (subsequently) 
relevant features $\feati{4}$ - $\feati{6}$.
\fref{fig:evaluation_synthetic} shows the relevant features for each packet of
the flow. Additionally, the dataset contains 4 features which are statistically
independent from the labels.

\myitem{CICIDS} The CICIDS2017 dataset~\cite{sharafaldinGeneratingNewIntrusion2018} consists of
network traffic during five days and contains various attacks. Each day contains
different attacks and the dataset comes with labels, which indicate the type
(malicious or benign) of each flow.

\myitem{UNIBS} The UNIBS-2009
dataset~\cite{UNIBS_web,dusiQuantifyingAccuracyGround2011,gringoliGTPickingTruth2009} consists of network
traffic from an edge router of the campus network of the University of Brescia
on three consecutive days in 2009. The dataset comes with labels which indicate
the protocol of each flow according to the DPI analysis by \texttt{l7filter}~\cite{l7filter}.

\startsquarepar
\myitem{Preprocessing} In the \emph{\cic dataset}, 
we aggregated "FTP-Patator" and "SSH-Patator"
into one attack type. 
In the \emph{\unibs dataset}, we ignored classes with less than $20$ samples.
Here, merging is not reasonable because traffic
classes represent different protocols.
\stopsquarepar

\startsquarepar
\myitem{Models} We used all stateful features from
\tref{tab:supported_features}. In all the models, the number of trees never
exceeded 32, and their depth never 20.
\stopsquarepar

\startsquarepar
\myitem{Simulation} \xref{subsec:evaluation_speed} and \xref{subsec:evaluation_accuracy}
rely on runtime simulations of the classification, written in Python. 
In both sections, we compare a full-accuracy classification with a classification as it 
would happen in hardware (notably, without floating-point operations). In order to achieve 
the latter, we translate $\RF s$ to their hardware counterpart, containing transformed 
integer thresholds with less accuracy~(cf.~\xref{subsec:compiler_memory}). We also 
simulate the full feature extraction in hardware, including error propagation due 
to reduced accuracy. 
We then track each flow across packet counts as its features pass through the $\RF s$, 
both for the full-accuracy and hardware mode. 
\stopsquarepar

\begin{table}[t]
        \small
        \renewcommand{\arraystretch}{1.2} 
        \begin{center}
        \begin{tabular}{@{}llrr@{}}
        \toprule
        \textbf{Dataset} & \textbf{Description} &  \textbf{PCAP Size} & \textbf{Flows} \\ 
        \midrule
        \multicolumn{4}{@{}l}{\textit{Synthetic}} \\
        \multicolumn{2}{@{}l}{\hspace{.2cm} Crafted to showcase context-dependent RF} & n/a$^a$ & 1\\ 
        \multicolumn{4}{@{}l}{\textit{CICIDS2017} \cite{sharafaldinGeneratingNewIntrusion2018}} \\
        \hspace{.2cm} Tue & FTP and SSH brute-force attacks & 7.8 GB & 282 418\\ 
        \hspace{.2cm} Fri~1& botnet& 8.3 GB & 130 353\\ 
        \hspace{.2cm} Fri~2& DDoS attacks& 8.3 GB & 114 836\\ 
        \hspace{.2cm} Fri~3& port scan& 8.3 GB & 248 837\\ 
        \multicolumn{4}{@{}l}{\textit{UNIBS-2009} \cite{UNIBS_web} } \\
        \hspace{.2cm} Day 1 & 8 application-layer protocols$^c$ & 318 MB$^b$ & 20 681\\ 
        \hspace{.2cm} Day 2 & 8 application-layer protocols$^c$ & 237 MB$^b$ & 19 657\\ 
        \hspace{.2cm} Day 3 & 6 application-layer protocols$^d$ & 2.0 GB$^b$ & 24 553\\ 
        \bottomrule
        \multicolumn{4}{@{}l}{\hspace{.2cm} $^a$ only feature values}\\
        \multicolumn{4}{@{}l}{\hspace{.2cm} $^b$ packets without payload}\\
        \multicolumn{4}{@{}l}{\hspace{.2cm} $^c$ bittorrent, edonkey, http, imap, pop3, skype, smtp, ssl} \\
        \multicolumn{4}{@{}l}{\hspace{.2cm} $^d$ bittorrent, edonkey, http, pop3, skype, ssl} \\
        \end{tabular}
        \caption{Evaluation datasets} 
        \label{tab:evaluation_datasets}
        \end{center}
    \end{table}

\subsection{Context-dependent random forests}\label{subsec:evaluation_synthetic}

\begin{figure}[b]
	\centering\includegraphics[width=0.9\linewidth]{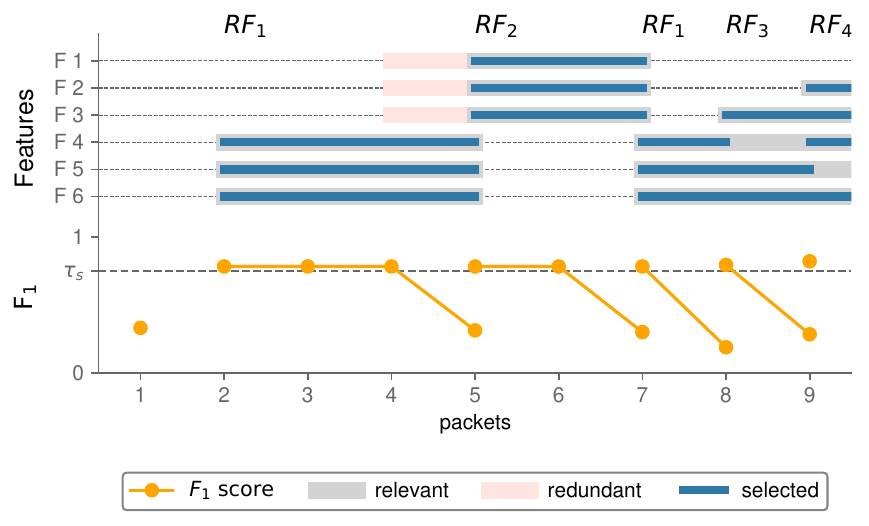}
	\caption{\system selects relevant features and switches to a new model if the score drops below $\boldsymbol{\thrS}$.}
	\label{fig:evaluation_synthetic}
\end{figure}

\startsquarepar
This experiment uses the synthetic dataset to show that \system successfully implements
the offline component of ASAP inference, \ie the training of context-dependent $\RF s$. 
The results in \fref{fig:evaluation_synthetic} indeed indicate that \system 
\first finds the first-possible $\RF$, 
\second recycles $\RF s$ as often as possible, and 
\third only selects relevant features.
\stopsquarepar

\myitem{First-possible RF use} \system successfully selects 
the first $\RF$ at packet count 2. This is the soonest 
a model can exceed the threshold score, according to the ASAP
inference design objective.

\newpage
\myitemNoVspace{Longest-possible RF reuse} \system applies 
a found $\RF$ as 
long as the features it uses are relevant and stable enough to 
allow for reusing the inferred thresholds -- \ie as long as 
its score is above $\thrS$. Iff the features change 
so much that $\RF_i$'s score is too low, \system generates 
a new $\RF$ (after 5, 7, 8 and 9 packets in
\fref{fig:evaluation_synthetic}). \system successfully switches to a previously found
$\RF$ instead of generating a new one, if a sufficiently good one 
is available (here, for packet 7).

\myitem{Locally minimal choice of necessary features} \system 
never chooses any of the irrelevant or redundant
features. 
Moreover, \system only uses the necessary \emph{subset} of
relevant features to achieve $\thrS$ (for instance, $\feati{2}$, $\feati{3}$,
$\feati{4}$, $\feati{6}$ for packet count 9, even though $\feati{5}$ would also be relevant). 

\startsquarepar
\smallskip\noindent
The efficacy of these mechanisms directly ties together with 
the optimization 
problem in \S\ref{subsec:rf_optimization} (objectives O1, O2, constraints C1, C2):
\stopsquarepar

First, \system optimizes for low memory consumption (O1): At runtime, 
applying the first $\RF$ as soon as possible also gives the earliest-possible chance 
for de-allocating per-flow state after classification. Recycling $\RF s$ as often as possible 
reduces the memory consumption of the overall $\RF$ sequence; and only selecting the relevant 
features brings down the per-flow state. 

\startsquarepar
Second, \system's training algorithm successfully contributes its part 
to maximizing classification speed (O2) by finding the first-possible $\RF$. 
In order to allow for per-flow speed tuning at runtime, it also 
extracts an $\RF$ sequence that can classify at later packet counts. 
\stopsquarepar

\startsquarepar
Third, the results inidicate that \system's training algorithm 
effectively constrains the model search with the threshold score $\tau_s$ (C1). 
\stopsquarepar

\startsquarepar
Fourth, by design, \system allows for setting model 
constraints like the maximum tree depth, and only relies on features that are 
implementable in \PPPP{} to guarantee feasibility in hardware (C2). 
\stopsquarepar

\subsection{Speed of ASAP inference at runtime}\label{subsec:evaluation_speed}

This experiment evaluates how quickly \system classifies
on a per-flow level, 
in terms of the packet count.
We run \system with the CICIDS and UNIBS dataset (cf.~Table~\ref{tab:evaluation_datasets}) and plot
the results in \fref{fig:evaluation_accuracy_speed}. 

The results show that \system classifies the majority of the flows within their
first few packets. In the CICIDS dataset, \system classifies 
99.1\perc of the flows (Tue), 
99.3\perc (Fri~1), 
89.3\perc (Fri~2), and 
82.5\perc (Fri~3)
after only 3 packets, with an F$_1$ score of 
99.6\perc (Tue), 
99.3\perc (Fri~1), 
99.9\perc (Fri~2), and 
97.9\perc (Fri~3). 
\system finishes classifying all flows at 12, 16, and 10 packets
with a total score of 99.6\perc, 99.5\perc, and 99.9\perc for 
Tue, Fri~1 and Fri~2, respectively. For Fri~3, a small percentage (1.8\perc)
of flows is not
classified up to a packet count of 45. In this case, the score for all classified flows
is 98.2\perc.

In the UNIBS dataset, \system classifies 
58.0\perc (Day~1), 83.0\perc (Day~2), and 66.9\perc (Day~3) 
after 5, 5, and 7
packets with an F$_1$ score of 
83.9\perc, 87.1\perc and 93.9\perc, respectively. 
The models end up classifying 
97.6\perc (Day~1), 98.9\perc (Day~2), and 98.6\perc (Day~3) 
of the flows at 24, 10, and 45 packets, 
with a score of 
84.8\perc, 86.4\perc and 84.1\perc, respectively. 
This also shows that a lower $\tau_s$ may result in less certain decisions occurring more often,
boosting the classification spread over several $\RF s$. 

All in all, for many of the considered datasets, 
one model would suffice to classify a large
share of the traffic, and \system indeed finds such a model.
However, if required, \system's context-dependent models classify traffic in
multiple stages. Day 1 of the UNIBS dataset shows such an example. There,
58.0\perc of the flows are classified after 5 packets, 82.7\perc after 7 packets,
and 91.2\perc after 10 packets.
\system therefore successfully defers uncertain decisions on a per-flow level to a later point, 
and indeed accepts labels when they are certain enough. 
This mechanism boosts the achieved overall classification score well above the threshold $\tau_s$.
We can therefore confirm the successful interplay of context-dependent $\RF s$ and 
certainty-based labeling, the two components of ASAP inference.

\subsection{Score of ASAP inference at runtime}\label{subsec:evaluation_accuracy}

\begin{figure}[t]
	\centering\includegraphics[width=0.9\linewidth]{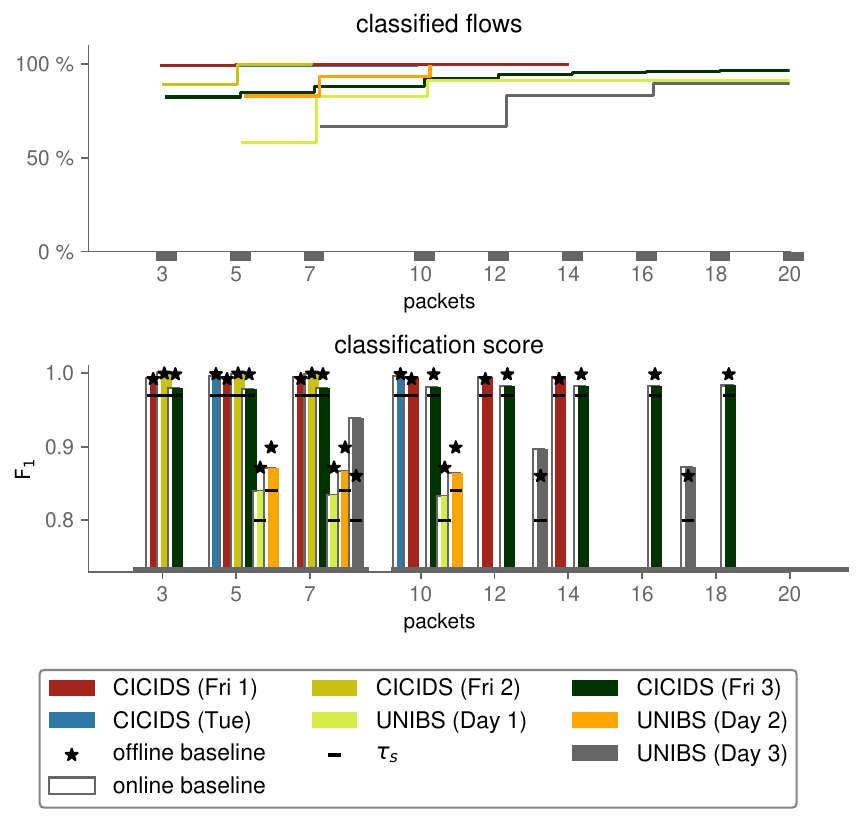}
	\caption{Percentage of classified flows and achieved accuracy. \system classifies a large percentage of the flows after few packets with a high score (F$_1$ macro score over all classes).}
	\label{fig:evaluation_accuracy_speed}
\end{figure}

Together with the classification speed (\cf above), we now report the achieved
score for these classifications.

In this experiment, we compare \system's F$_1$ macro score over all traffic
classes with an offline and an online baseline. The \emph{online baseline} shows
the case where the same models that \system applies in the data plane are
applied in software with floating point operations (while \system reduces the
precision of features in order to save memory and does not have floating point
operations available, \cf \xref{sec:compilation_to_p4}). The \emph{offline baseline} shows
the case where the full flows are classified by a model that is trained on full
flows (\ie no \emph{ASAP} classification) and with all features. 

\fref{fig:evaluation_accuracy_speed} visualizes the results of this experiment.
We \first compare with $\thrS$ and the two baselines, \second discuss a general 
approximation of the baselines, and \third analyze the effect of $\thrC$. 

\myitem{Comparison with the baselines and threshold score}
We first note that \system achieves a high score which is on-par with models
running in software. \system is never more than 0.03\perc below the online
baseline and never more than 7.9\perc below the offline baseline.
\system exceeds the respective $\thrS$ in all cases.
For the CICIDS dataset, we observe that \system is very close to the offline
baseline (from 0.2\perc for Fri~2 up to 2\perc for Fri~3).
For the CICIDS Tue, Fri~3, and UNIBS Day~1 and Day~2 datasets, there is
zero difference between \system's inference and the online baseline; in these
cases, a perfect translation was possible.
For the UNIBS Day~3 dataset, \system momentarily exceeds the offline baseline, 
showing
that classifying subflows can actually improve the score. One possible
explanation is that this is purely due to the certainty threshold, \ie
the fact that the online version only classifies a subset of flows while
the offline version classifies all flows.
Another possible
factor could be that flows for different applications might show
characteristic behavior mainly at the beginning, and that this 
signature fades out over time.

\myitem{Arbitrary approximation of both baselines} 
First, \system could approximate the offline baseline arbitrarily 
well: If one selects $\tau_s$ to be equal the offline baseline, one is guaranteed
to find at the latest the offline model in the training process. 
Second, \system can also approximate the online baseline by increasing the parameter
for the translation precision (\cf \S\ref{subsec:compiler_memory}). 
Trivially, this can incur a cost on the number of flows that can be stored concurrently, and on 
the classification speed. This highlights the trade-off of competing objectives
that we laid out in the optimization problem~(\S\ref{subsec:rf_optimization}), namely a trade-off
between classification score, memory usage, and speed. 

\startsquarepar
\myitem{Effect of $\boldsymbol{\tau_c}$} The results show that the certainty threshold allows
to exceed the $\thrS$  by far (\eg 14\perc for UNIBS Day 3, at packet count 7). 
Simultaneously, the certainty threshold allows
to tune the percentage of classified flows. If $\thrC$ was $0$, the 
overall performance would barely exceed $\thrS$ since all flows 
(that fit into memory) would be 
classified with the first $\RF$ in the sequence.
Hence,
each flow would occupy the memory for as little time as possibly achievable.
Hence, varying $\thrC$  over time is analogous 
to a control system on memory usage. 
\system allows for an easy implementation of such a system via
communication with the controller, since $\thrC$ is adaptable at runtime. 
Such an extension lies beyond the scope of this paper.
\stopsquarepar

\vspace{-0.04cm}
\subsection{Memory consumption of ASAP inference}\label{subsec:evaluation_scalability}

\begin{figure}[t]
    \captionsetup{belowskip=5pt}
	\centering\includegraphics[width=0.9\linewidth]{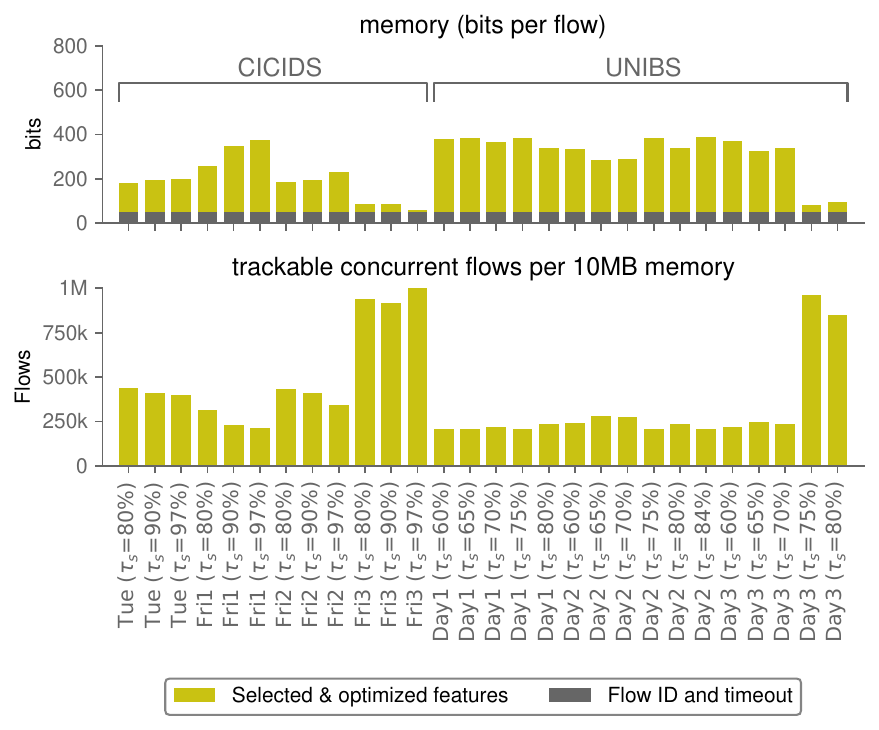}
    \caption{Number of bits required to store per-flow features and number of
    trackable flows per 10MB of memory. \system's feature selection
    and optimization allows for tracking hundreds of
    thousands of flows per 10MB available memory.}
	\label{fig:evaluation_memory}
    \vspace{-0.33cm}
\end{figure}

\startsquarepar
In this experiment, we evaluate the amount of memory that \system requires per
flow in order to store its features, in dependency of different $\thrS$ 
(\fref{fig:evaluation_memory}). We again evaluate the 
UNIBS and CICIDS dataset.
\stopsquarepar

The amount of per-flow memory consists of two parts: \first a model-independent
part for the flow ID and timestamp, requiring in 49 bits; and \second
a model-dependent part for the feature storage.

\startsquarepar
The absolute number of bits per flow depends on the dataset and
$\thrS$. This is expected because depending on the dataset and
$\thrS$, \system requires a different number and complexity of models.
\stopsquarepar

\startsquarepar
Since a higher $\thrS$ requires better models, the intuition is that 
this results in more features, and thus a higher memory consumption. 
Indeed, this is what we observe for \eg the CICIDS~Fri~1 dataset.
However, interestingly, this is not the case for \eg UNIBS~Day~3. 
\stopsquarepar

We find that the main reason for this effect is the following: $\thrS$ can be so high that one finds fewer $\RF s$ that satisfy this requirement. 
For instance, for UNIBS~Day~3, $\thrS = 0.6$ results in eight $\RF s$ at packet counts
$3, 5, 7, 10, \ldots$ , but $\thrS = 0.8$ results
in only four $\RF s$ that are applicable at packet counts $5, 7, 10, 22$, and $24$.

Note that since \system relies on floating-point features, it is not possible to compare
our allocation strategy with a ``default'' strategy for selecting the number of 
significant digits. Since \system allocates the features based on the feature
thresholds in the $\RF s$, it is also not possible to estimate how much space 
non-selected features would consume if one stored them ``just-in-case''.

%% file: sec_discussion.tex
\pagebreak
\section{Discussion} \label{sec:discussion}

In this section, we discuss important properties, limitations and possible
extensions of \system.

\myitem{Classifying other entities}
In its current form, \system classifies flows. However, the same approach works
for classifying other entities (\eg individual packets or hosts). 

\myitem{Resource exhaustion attacks} 
Because of the limited memory, \system can only track a fixed number of
concurrent flows. A malicious actor could initiate many different flows in order
to exhaust the available memory. However, \system can detect idle flows based on
the timestamp of their last packet and it reuses their memory cells for another
flow.

\myitem{Distributing the classification}
\system implements a simple best-effort strategy to flag classified flows (\cf
\xref{sec:inference}). A more sophisticated -- and feasible -- approach
would be to
actively route flows via devices that have capacity to classify them. 

\myitem{Other datasets}
The absolute performance of \system in terms of score and memory depends on
the training dataset, but \system computes models and feature sets that maximize
the score and minimize memory requirements.
Datasets can be obtained from a third party (\eg
\cite{IDS2017D62:online,ADFAIDSDatasets2013}) or recorded from the own network
(with automated labeling \cite{lizhiTrafficLabellerCollecting2014a, gringoliGTPickingTruth2009}).

\startsquarepar
\myitem{Different packet sampling strategies} 
Currently, \system's training algorithm creates an $\RF$ sequence over sampled packet counts
(at 1, 3, 5, \ldots packets). This is just one example for a possible sampling granularity. 
Generally, the optimal sampling strategy depends on 
the available memory (since the number of $\RF s$ correlates with the sampling density); 
the priority of classification speed; 
the expected variation of feature values over packet counts; 
constraints on the total training time; and
the timing of the most relevant patterns for classification. 
A sampling strategy that takes all of these factors into account is an interesting avenue
for future work. 
\stopsquarepar

%% file: sec_related.tex
\newpage
\section{Related work} \label{sec:related}

This section compares \system to closely related approaches. 
We start discussing approaches which also aim at performing inference with 
$\RF s$ or $DTs$ on network switches before discussing other in-network 
machine learning approaches.

\myitem{$\boldsymbol{RFs}$ and DTs on network switches}
Zheng \etal and Xiong \etal have proposed several systems for $\RF s$
and $DTs$ in the data 
plane~\cite{xiongSwitchesDreamMachine2019,zhengPlanterSeedingTrees2021,zhengIIsyPracticalInNetwork2022,zhengAutomatingInNetworkMachine2022}, 
the most recent iteration being Ilsy~\cite{zhengIIsyPracticalInNetwork2022} and 
Planter~\cite{zhengAutomatingInNetworkMachine2022}. None of these systems optimizes for early classification, unlike \system. Indeed, Ilsy trains a single model and translates it to the data plane. Ilsy's translation differs from \system in that it partitions the feature space according to the subsequent trees' decisions, and uses exact-match tables to map a flow's feature values to the corresponding leaf node. This translation has pros and cons. On the one hand, Ilsy's translation does not depend on the depth of the trees and, as such, it can theoretically support deeper models than \system. On the other hand, Ilsy's translation leads to large exact-match tables for features values that are wide-ranging. In the datasets we used for instance, some features (\eg inter-arrival times) required more than 40 bits in the data plane.

\startsquarepar
Planter translates already-trained models and improves Ilsy's 
translation strategy: It uses range-match instead of exact-match tables. This 
means that the feature table size now scales with the number of distinct feature thresholds used in the $\RF$. 
However, range-match tables are located in TCAM, which is far more limited than the SRAM for exact-match tables. 
Since \system generates an entire sequence of $\RF s$, the number of thresholds may exceed what 
Planter expects to encounter: For instance, the packet length average feature for the CICIDS 
Tuesday dataset gets compared with a total of 5942 distinct thresholds. In contrast to Planter's 
translation strategy, \system directly stores the $\RF$'s thresholds and compares the features explicitly. 
All in all, the best compilation strategy depends on the used features, the required model depth, 
and the number of thresholds per feature. It would be interesting 
to automatically select the best method on a case-by-case basis in future work.
\stopsquarepar

Mousika~\cite{xieMousikaEnableGeneral2022} and~\citep{koganExploitingOrderIndependence2016} 
compress already-trained models such that they fit into the data plane, but without optimizing for early classification, unlike \system. That said, their compression techniques might help compressing \system's models further. Given its loss-freeness, \cite{koganExploitingOrderIndependence2016} compression might work out of the box on \system model; this is not the case for Mousika though as it currently does not support stateful features.

FlowLens~\cite{barradasFlowLensEnablingEfficient2021a} generates memory-efficient
feature representations by aggregating them in bins and only using the most relevant bins. 
However, the bins' value range corresponds to powers of 2 due to hardware restrictions. 
Hence, the ranges may be far off the feature threshold domains that \system's $\RF s$ use, 
and therefore incur a high penalty in classification accuracy. 

Homunculus~\cite{swamyHomunculusAutoGeneratingEfficient2022}, pHeavy~\cite{zhangPHeavyPredictingHeavy2021},
and SwitchTree~\cite{leeSwitchTreeInnetworkComputing2020} propose methods for training and 
translating $DTs$ or $\RF s$ to the data plane. 

\newpage
Homunculus~\cite{swamyHomunculusAutoGeneratingEfficient2022} generates machine learning models
according to hardware constraints, and translates them directly onto the switch. 
The authors showcase Homunculus with $\RF s$. For the training, they use Bayesian 
optimization to explore the design space -- however, they do not support generating 
a sequence of machine learning models. For the translation to \ma table platforms, 
they use Ilsy (see above). In contrast, \system fully explores a non-sampled grid of eligible model parameters for each $\RF$ meaning \system solutions should be least as good as those found by Homunculus. In the future though, one could speed up the training process by relying on Homunculus’ strategy.

pHeavy~\cite{zhangPHeavyPredictingHeavy2021} spreads out the detection of heavy flows over several 
models. However, in order to confirm a flow is heavy, one needs to always apply the entire sequence.
Since the models in the sequence specialize on different subsets of the training dataset, pHeavy
does \emph{not} attempt to reapply models for as long as possible, or to reuse them. In contrast, 
\system attempts classification for \emph{all} labels with each $\RF$, and defers the decision \emph{only if} it is uncertain. 
 
SwitchTree~\cite{leeSwitchTreeInnetworkComputing2020} and \cite{xavierProgrammableSwitchesInNetworking2021}
train and implement $\RF s$ or $DTs$, respectively, in switches. Unlike \system. none of them not support early classification. 
Moreover, SwitchTree does not support automatic model parameter tuning.

NetPixel~\cite{siddiqueNetworkacceleratedMLbasedDistributed2021} uses an in-switch $DT$ to classify images. Contrary to \system, it does not classify flows, but instead performs inference on the sent data using a custom protocol. 

\myitem{Other in-network machine learning}
Apart from $\RF s$, there has been work to implement 
neural networks in NICs~\cite{sanvitoCanNetworkBe2018,siracusanoDeepLearningInference2018,siracusanoRearchitectingTrafficAnalysis2022},
switching chips~\cite{siracusanoInnetworkNeuralNetworks2018,qinLineSpeedScalableIntrusion2020a,saquettiInNetworkIntelligenceRunning2021},
or even switches enhanced with optical matrix multiplication~\cite{zhongIOIInnetworkOptical2021}; 
to translate clustering algorithms to switches~\cite{friedmanClustreamsDataPlane2021a,bremler-barrUltraFastSimilaritySearch2015} 
and to use reinforcement learning in NICs~\cite{simpsonRevisitingClassicsOnline2022}. 
Since the best machine learning choice depends on the concrete task at hand, we view our work
as complementary to these approaches. Indeed, $\RF s$ have been successfully used for 
traffic classification~\cite{arzaniPrivateEyeScalablePrivacyPreserving2020,arzaniTakingBlameGame2016,hayesKfingerprintingRobustScalable2016}.
Moreover, $\RF s$ can be used for clustering~\cite{shiUnsupervisedLearningRandom2006} 
or to approximate neural networks~\cite{frosstDistillingNeuralNetwork2017}.

%% file: sec_conclusion.tex
\section{Conclusion} \label{sec:conclusion}

\startsquarepar
We presented \system, a novel framework for performing ASAP in-network inference
using $\RF s$. Unlike existing in-network classifiers, \system applies
the classification after having seen the smallest amount of packets possible, and while still guaranteeing a high score. The key technical insight behind \system is to train a sequence of $\RF s$, and to automatically switch
from one to the other as the flow progresses. \system manages to optimize these models at training time according to the constraints of programmable data planes allowing them to run in existing hardware. As we show, this can be done without lowering the achieved classification score.
\stopsquarepar

\startsquarepar
We implemented a prototype of \system and showcased its practicality. 
Our evaluation shows that \system manages to perform accurate ASAP classification for 
hundreds of thousands of flows.
\stopsquarepar

%% file: app_training_alg.tex
\onecolumn
\section{Training algorithm}\label{sec:app_training_alg}

The pseudocode below describes the \system greedy algorithm
\pxref{subsec:rf_training}.
The Python code for training consists of $\approx 17000$ lines
of code, the P4 code for inference consists of $\approx 1500$ lines of code.

\begin{algorithm}
    \DontPrintSemicolon

    \KwIn{         
		$\mathtt{P}$, packet counts to analyse \hfill \break
		$\featurelist = [\ldots,\features{\Flowset[:p]}, \ldots], p \in P$, List of features for up to p packets \hfill \break 
    		$\features{\Flowset}$, Features of completed flows \hfill \break
    		$\mathtt{thr}$, F$_1$ Threshold \hfill \break
    		$\rflabel{\Flowset}$, True labels \hfill \break
    		$T$, initial trade-off parameters to select representative features
    		
    }
    \vspace{.2cm}
    \KwOut{        
    		$\mathtt{M}$, List of at which packet count to use what model with what 
features
    		
    }
    \vspace{.2cm}
    
\tcp*{----- find redundant groups of features -----}	
$\midistance \gets \textsc{mi\_distance}(\features\Flowset)$ \tcp*{information 
distance between features}
	$\mathtt{G} \gets \textsc{dbscan}(distance\_matrix = \midistance)$ 
\tcp*{redundant feature groups} 
	$\mathtt{M} \gets []$ \tcp*{init extracted models}
    \While{$\featurelist$ not empty} 
    { 
		\tcp*{----- model search -----}   
		$s \gets 0$ \tcp*{init score with 0}
        \While{$\featurelist$ not empty AND $s \leq \mathtt{thr}$} {
        		$p \gets \mathtt{P}.\text{pop}(0)$ \tcp*{get next packet count}
        		$f \gets \featurelist.\text{pop}(0)$ \tcp*{get next features}
        		$T \gets \textsc{updateTradeOff}(T, p)$

        		\tcp*{----- selecting representative features -----}
        		$f_r \gets \textsc{representative}(\mathtt{G}, T)$ 
        		
        		$(\RF, s) \gets \textsc{gridsearch}(r, \rflabel{\Flowset})$ 
\tcp*{search for optimal RF and score}
        }
        \If{$\featurelist$ is empty} {
        		break
        }
        \tcp*{----- model optimization -----}
        $f_s \gets \textsc{SelectWithScore}(\RF, f_r, \mathtt{thr})$ 
\tcp*{select minimal features and retrain \RF}
        $f_{names} \gets f_s.\text{names}$ \tcp*{get best feature names}
        $T \gets \textsc{updateUtility}(T, f_{names})$ \tcp*{increase likelihood 
of reselection}
        $\RF_p \gets \RF$
        
        $\mathtt{M}.\text{append}([p, \RF_p, f_{names}])$

        \tcp*{----- longest-possible model reapplication -----}
        \While{$\featurelist$ not empty AND $s > \mathtt{thr}$} {
        		$p \gets \mathtt{P}.\text{pop}(0)$ \tcp*{get next packet count}
        		$f \gets \featurelist.\text{pop}(0)$ \tcp*{get next features}
        		$f_s \gets f.\text{keep}(f_{names})$ \tcp*{keep best features}
        		
        		$s \gets \textsc{score}(\RF_p, f_s, \rflabel{\Flowset})$
        		
        		\If{$s \leq \mathtt{thr}$}{
				$\RF_{old}, f_{names\_old}, s_M \gets \textsc{BestOldRF}(\mathtt{M})$
				
				\If{$s_M > \mathtt{thr}$} { 
					$\RF_p, f_{names} \gets (\RF_{old}, f_{names\_old})$ \tcp*{reuse old \RF}
					$\mathtt{M}.\text{append}([p, \RF_p, f_{names}])$ 
				}
				\Else{
					$\mathtt{P}.\text{insert}(0, p)$ \tcp*{reinsert current count at index 0}
					$\featurelist.\text{insert}(0, f)$ \tcp*{reinsert current features at index 
0}				
				} 
%
%
        		}
        }
    }



	\vspace{.3cm}
	\caption{The \system greedy algorithm}
	\label{alg:training}
\end{algorithm}